\documentclass[
    aps,
    prd,
    notitlepage,
    nofootinbib,
    10pt,
    oneside,
    a4paper,
    preprintnumbers,
    superscriptaddress,
    longbibliography,
    final,
]{revtex4-2}

\usepackage{silence}
\usepackage[utf8]{inputenc}
\usepackage[english]{babel}
\usepackage{microtype}

\usepackage{lmodern}
\usepackage[margin=2.5cm]{geometry}

\usepackage{amsmath,mathtools,amssymb}
\usepackage{wasysym} %
\usepackage{bm}
\usepackage{tensor}
\usepackage{graphicx}

\usepackage[svgnames]{xcolor}
\usepackage{booktabs, multirow}
\usepackage[allcolors=black]{hyperref}
\usepackage{cleveref,csquotes}
\crefname{equation}{equation}{equations}
\AddToHook{cmd/appendix/before}{\crefalias{section}{appendix}}
\usepackage[
    printonlyreused,
    dua, %
]{acronym}
\usepackage[
    free-standing-units,
    exponent-product = \cdot,
    list-units = single,
    range-units = single,
    multi-part-units = single,
    separate-uncertainty = true,
]{siunitx}
\usepackage{float}
\usepackage{physics2}
\usephysicsmodule{ab} %
\usepackage{xfrac}
\usepackage{iftex} %

\usepackage{tikz}
\usepackage{pgfplots}

\usetikzlibrary{
    external,
    perspective,
    calc,
    arrows.meta,
}
\ifLuaTeX
\else
\fi
\usepgfplotslibrary{
    groupplots,
    external,
}
\pgfplotsset{
    scale only axis,
    width=0.5\textwidth,
    height=0.2\textheight,
    compat = newest,
    every axis plot/.append style={
        line join=round,
        line cap=round,
        clip=false,
    },
    x axis line style = {thick},
    y axis line style = {thick},
    legend style={draw=none,fill=none,},
    legend style={/tikz/every even column/.append style={column sep=0.3cm}},
}

\ifLuaTeX %
    \usepackage{luacode}
    \begin{luacode*}
        -- Linear algebra
        linalg = require("lua.linalg")
        Vec3, Mat3x3 = linalg.Vec3, linalg.Mat3x3

        -- Plotting angles
        tex_cmds = require("lua.tex_cmds")
        arc_segments = tex_cmds.arc_segments
        draw_arc_gradient = tex_cmds.draw_arc_gradient
        draw_line_gradient = tex_cmds.draw_line_gradient
        right_angle_mark_segments = tex_cmds.right_angle_mark_segments

        -- Trigonometry
        sin, cos = math.sin, math.cos
        rad, deg = math.rad, math.deg

        -- Shorthand
        function tp2s(txt) tex.print(tostring(txt)) end

        function make_angle_positive(angle)
            if angle >= 0 then return angle
            else return make_angle_positive(2 * math.pi + angle) end
        end

        -- Physical constants
        G_OVER_C_SQUARE = 1
    \end{luacode*}
\fi

\DeclareMathOperator{\atantwo}{atan2}

\newcommand{\ie}{\textit{i.e}}
\newcommand{\eg}{\textit{e.g}}

\newcommand{\bigO}{\mathcal{O}}
\newcommand{\bigOof}[1]{\bigO\left(#1\right)}
\newcommand{\norm}[1]{\left\lVert#1\right\rVert}

\renewcommand{\vec}[1]{\bm{#1}}
\newcommand{\uvec}[1]{\hat{\bm{{#1}}}}
\makeatletter
\newcommand\vu[1]{\@ifstar{\uvec{#1}}{\uvec{#1}}}
\makeatother

\newcommand{\thetaE}{\theta_\text{E}}
\newcommand{\RE}{R_\text{E}}

\newcommand{\RS}{R_\text{Sch}}
\newcommand{\openangle}{\gamma}
\newcommand{\openanglep}{\gamma_+}
\newcommand{\openanglem}{\gamma_-}
\newcommand{\openanglepm}{\gamma_\pm}
\newcommand{\MSMBH}{M_\text{SMBH}}
\newcommand{\Mbin}{M_\text{bin}}
\newcommand{\rorbit}{r_\text{orbit}}
\newcommand{\vorbit}{v_\text{orbit}}
\newcommand{\vorbitp}{v_{+}}
\newcommand{\vorbitm}{v_{-}}
\newcommand{\vorbitpm}{v_{\pm}}
\newcommand{\vorbitvec}{\vec{v}_\text{orbit}}

\newcommand{\Nhat}{\uvec{k}}
\newcommand{\Nhatp}{\uvec{k}_+}
\newcommand{\Nhatm}{\uvec{k}_-}
\newcommand{\Nhatpm}{\uvec{k}_\pm}
\newcommand{\robs}{r_\text{obs}}

\newcommand{\robsvec}{\vec{r}_\text{obs}}
\newcommand{\rsrcvec}{\vec{r}_\text{src}}
\newcommand{\rlnsvec}{\vec{r}_\text{lns}}

\newcommand{\ascnodehat}{\uvec{\ascnode}}
\newcommand{\ascnodehatp}{\uvec{\ascnode}_+}
\newcommand{\ascnodehatm}{\uvec{\ascnode}_-}
\newcommand{\ascnodehatpm}{\uvec{\ascnode}_\pm}

\newcommand{\Lperphat}{\uvec{L}_\perp}

\newcommand{\Lperphatpm}{\uvec{L}_{\perp,\pm}}

\newcommand{\ra}{\alpha} %
\newcommand{\dec}{\delta} %

\newcommand{\incl}{\iota}
\newcommand{\inclp}{\iota_+}
\newcommand{\inclm}{\iota_-}
\newcommand{\inclpm}{\iota_\pm}

\newcommand{\thetaJN}{\theta_{JN}}
\newcommand{\thetaJNp}{\theta_{JN,+}}
\newcommand{\thetaJNm}{\theta_{JN,-}}
\newcommand{\thetaJNpm}{\theta_{JN,\pm}}

\newcommand{\thetaLN}{\theta_{LN}}
\newcommand{\thetaJL}{\theta_{JL}}

\newcommand{\phiJL}{\phi_{JL}}
\newcommand{\phiJLp}{\phi_{JL,+}}
\newcommand{\phiJLm}{\phi_{JL,-}}
\newcommand{\phiJLpm}{\phi_{JL,\pm}}

\newcommand{\philns}{\phi_\text{lns}}
\newcommand{\thetalns}{\theta_\text{lns}}

\newcommand{\phiorb}{\Phi}
\newcommand{\phiorbp}{\Phi_{+}}
\newcommand{\phiorbm}{\Phi_{-}}
\newcommand{\phiorbpm}{\Phi_{\pm}}

\newcommand{\tcoal}{t_\text{coal}}

\newcommand{\phase}{\varphi}

\newcommand{\phicoal}{\phi_\text{coal}}

\newcommand{\phicoalpm}{\phi_{\text{coal},\pm}}

\newcommand{\psip}{\psi_+}
\newcommand{\psim}{\psi_-}
\newcommand{\psipm}{\psi_\pm}

\newcommand{\phimis}{\phi_\text{mis}}
\newcommand{\thetamis}{\theta_\text{mis}}

\newcommand{\Lvec}{\vec{L}_\text{src}}
\newcommand{\Jvec}{\vec{J}_\text{src}}
\newcommand{\Lhat}{\uvec{J}_\text{src}}
\newcommand{\Jhat}{\uvec{J}_\text{src}}
\newcommand{\LAGNvec}{\vec{L}_\text{AGN}}
\newcommand{\LAGNhat}{\uvec{L}_\text{AGN}}

\newcommand{\Dl}{D_\ell}
\newcommand{\Dls}{D_{\ell s}}
\newcommand{\Ds}{D_s}

\newcommand{\dL}{d_\text{L}} %

\newcommand{\zhat}{\uvec{z}}

\newcommand{\xhatL}{\uvec{x}_\text{L}}
\newcommand{\yhatL}{\uvec{y}_\text{L}}
\newcommand{\zhatL}{\uvec{z}_\text{L}}

\newcommand{\xhatJ}{\uvec{x}_\text{J}}
\newcommand{\yhatJ}{\uvec{y}_\text{J}}
\newcommand{\zhatJ}{\uvec{z}_\text{J}}

\newcommand{\xhatJp}{\uvec{x}_{\text{J},+}}
\newcommand{\yhatJp}{\uvec{y}_{\text{J},+}}

\newcommand{\xhatJm}{\uvec{x}_{\text{J},-}}
\newcommand{\yhatJm}{\uvec{y}_{\text{J},-}}

\newcommand{\xhatJpm}{\uvec{x}_{\text{J},\pm}}
\newcommand{\yhatJpm}{\uvec{y}_{\text{J},\pm}}
\newcommand{\zhatJpm}{\uvec{z}_{\text{J},\pm}}

\newcommand{\xhatlns}{\uvec{x}_\ell}
\newcommand{\yhatlns}{\uvec{y}_\ell}
\newcommand{\zhatlns}{\uvec{z}_\ell}

\newcommand{\denomInJFrame}{\Delta_\text{J}}

\newcommand{\dd}{\mathrm{d}}

\newcommand{\defdby}{\coloneqq}

\newcommand{\signalduration}{\tau_\text{sig}}

\acrodef{gr}[GR]{general relativity}
\acrodef{gw}[GW]{gravitational wave}
\acrodef{bh}[BH]{black hole}
\acrodef{bbh}[BBH]{binary black hole}
\acrodef{cbc}[CBC]{compact binary coalescence}
\acrodef{ns}[NS]{neutron star}
\acrodef{bns}[BNS]{binary neutron star}
\acrodef{pbh}[PBH]{primordial black hole}
\acrodef{agn}[AGN]{active galactic nucleus}
\acrodefplural{agn}[AGNs]{active galactic nuclei}
\acrodef{em}[EM]{electromagnetic}
\acrodef{smbh}[SMBH]{supermassive black hole}
\acrodef{cc}[c.c.]{cosmological constant}
\acrodef{hl}[HL]{Ho\v{r}ava-Lifshitz}
\acrodef{qnm}[QNM]{quasinormal modes}
\acrodef{lvk}[LVK]{LIGO-Virgo-KAGRA}

\makeatletter
\AtBeginDocument
 {
   \def\ltx@label#1{\cref@label{#1}}%
   \def\label@in@display@noarg#1{\cref@old@label@in@display{#1}}%
\def\label@in@mmeasure@noarg#1{%
    \begingroup%
      \measuring@false%
      \cref@old@label@in@display{#1}%
    \endgroup}%
 } %
\makeatother

\DeclareSIUnit\solarmass{M\textsubscript{$\odot$}}
\DeclareSIUnit\RSunit{\ensuremath{\mathit{R}_{Sch}}}
\DeclareSIUnit\parsec{pc}

\definecolor{ggTableauTenOne}{HTML}{4e79a7} %
\definecolor{ggTableauTenTwo}{HTML}{f28e2b} %
\definecolor{ggTableauTenThree}{HTML}{e15759} %
\definecolor{ggTableauTenFour}{HTML}{76b7b2} %
\definecolor{ggTableauTenFive}{HTML}{59a14f} %
\definecolor{ggTableauTenSix}{HTML}{edc948} %
\definecolor{ggTableauTenSeven}{HTML}{b07aa1} %
\definecolor{ggTableauTenEight}{HTML}{ff9da7} %

\definecolor{LvecCol}{RGB}{142,88,178} %
\definecolor{JvecCol}{RGB}{225,128,14} %
\definecolor{WaveCol}{HTML}{2CAB76} %
\definecolor{LnsPlaneCol}{HTML}{edc948} %
\definecolor{PlusCol}{HTML}{FF6347} %
\definecolor{MinusCol}{HTML}{6495ED} %

\definecolor{RedShiftCol}{RGB}{192, 68, 131} %
\definecolor{BlueShiftCol}{RGB}{58, 133, 166} %

\crefname{section}{section}{sections}
\crefname{subsection}{subsection}{subsections}
\crefname{figure}{figure}{figures}

\newcommand{\imagecolor}{Tomato}
\newcommand{\imagecolordn}{CornflowerBlue}

\begin{document}

\preprint{}
\title{
    Stereoscopic gravitational-wave probe: \texorpdfstring{\\}{}
    Gravitational lensing of a binary merger in active galactic nuclei
}

\author{Paul Martens}
\email{paulmartens@cuhk.edu.hk}
\affiliation{Department of Physics, The Chinese University of Hong Kong (CUHK), Shatin, New Territories, Hong Kong}

\author{Samson H. W. Leong}
\email{samson.leong@link.cuhk.edu.hk}
\affiliation{Department of Physics, The Chinese University of Hong Kong (CUHK), Shatin, New Territories, Hong Kong}

\author{Ryan Zhang}
\email{yzhan629@jhu.edu}
\affiliation{William H. Miller III Department of Physics and Astronomy, Johns Hopkins University, 3400 North Charles Street, Baltimore, Maryland, 21218, USA}

\author{Otto A. Hannuksela}
\email{hannuksela@phy.cuhk.edu.hk}
\affiliation{Department of Physics, The Chinese University of Hong Kong (CUHK), Shatin, New Territories, Hong Kong}

\date{\today}

\begin{abstract}
    \noindent
    Recent work has shown that, if a large fraction of the LIGO-Virgo-KAGRA compact binary coalescences (CBCs) occur in accretion disks around active galactic nuclei (AGNs), many such AGN-CBC systems will likely experience lensing and produce multiple gravitational-wave \enquote{images} of the CBCs that will become detectable with current or future gravitational-wave detectors.
    However, the waveforms produced by such lensed \enquote{images} will differ from those produced by other lensing systems because the binary source is orbiting the AGN's central supermassive black hole at a close distance.
    Here, we show that these AGN lensing systems produce two snapshots of the same binary source from two different viewing angles.
    These differences in the effective viewing angle between the images are encoded in the gravitational waves they emit, such that gravitational-wave detectors can effectively analyse the waveforms from two viewpoints.
    Beyond such \enquote{multi-view lensing}, the gravitational-wave images may also differ by their Doppler shift due to the expected relativistic orbital motion of the source around the AGN, an effect which is further amplified by the proximity to the lens.
    If the AGN's own angular momentum can be related to the binary's orbital angular momentum, we can describe the entire AGN-CBC system using only one additional free parameter compared with a typical point-mass lens system, or with three additional parameters when we do not make such an assumption.
    In this work, we establish a framework of transformation rules that describe both the angles in the two points of view, and the Doppler effect that comes from the source's motion around the lens.
    We also provide the framework to produce consistent AGN-lensed waveforms from any currently available gravitational-wave waveform.
    These AGN-CBC systems give two unique points of view on a binary merger, akin to a camera taking snapshots from two viewing angles, thus opening up a new window into the merger's properties and drawing the first steps on a path to understanding a lensing scenario that is fundamentally unlike any other lens system.
\end{abstract}

\maketitle

\section{Introduction}

With the formulation of general relativity came the famous three tests of general relativity: the perihelion precession of Mercury, the bending of light and the gravitational redshift.
A fourth one could arguably also have been added: gravitational waves.
At the time, however, they were expected to be so tiny that they would never be observed \cite{einsteinApproximativeIntegrationField1916,einsteinUberGravitationswellen1918}.
Yet, in 2015, the LIGO-Virgo collaboration observed gravitational waves for the first time, thus opening a completely new window on the Universe \cite{LIGOScientific:2016aoc}, confirming the predictive power of general relativity in the strong-field limit, and discovering a new population of stellar-mass black holes.
Since then, hundreds of signals have been observed and confidently associated with binary mergers~\cite{LIGOScientific:2026wfs}.

Beyond such famous confirmations of the general-relativistic predictions, another one postulates that gravitational waves, like light, will be gravitationally lensed as they propagate near massive objects.
While no such occurrence has yet been observed~\cite{LIGOScientific:2025cwb,LIGOScientific:2023bwz,LIGOScientific:2021izm}, confident detections of lensed gravitational-wave events are expected in the near future \cite{Ng:2017yiu,Haris:2018vmn,Oguri:2018muv,Li:2018prc,Smith:2022vbp,ET:2019dnz,Evans:2023euw}.
Gravitational lensing usually invokes three entities: a source, an observer, and a lens, placed between the former two.
Typical textbook introductions of the subject first treat the simple case of a point-mass lens, which describes the lensing of a source by a point-like object.
Such point-like lenses are a good first-order approximation to describe lensing by black holes.

\bigskip

Here we consider the lensing of gravitational waves emitted by a binary merger trapped inside the accretion disk of an \ac{agn} and orbiting its central \ac{smbh}.
Such systems are known in the literature as \enquote{b-EMRI}, for binary (b-) in an extreme mass ratio inspiral (EMRI), or sometimes as \enquote{hierarchical triple system}, when the nature of the three massive bodies is more arbitrary (\eg., stars, black holes,...).
In the present case, we denote as \enquote{AGN-CBC} the specific case where the \ac{smbh} is the central black hole of an \ac{agn}.
This denomination underlines that some extra assumptions can be made on the system (\eg., presence of an accretion disk, ...).
Independently of lensing, such systems are currently gaining interest for various reasons \cite{Samsing:2020tda,Sedda:2023big,Morton:2023wxg,Leong:2024nnx,McKernan:2012rf,McKernan:2014oxa,McKernan:2023xio,Bartos:2016dgn,Bellovary:2015ifg,Peng:2021vzr,Li:2025iux}.
For one, an AGN-CBC represents a possible formation channel for binaries of \acp{ns} or black holes \cite{Yang:2019cbr,Samsing:2020tda,Sedda:2023big,Mapelli:2020vfa}.
At its core, the idea is for the \ac{agn} to bring compact objects together into binaries in its vicinity.
More specifically, they would gather and merge in so-called \enquote{migration traps} \cite{Thompson:2005mf,Bellovary:2015ifg,Secunda:2018kar,Peng:2021vzr,Graham:2020gwr,Leong:2024nnx}, specific distances in the accretion disk where the changing torque direction of the disk brings orbiting objects together.
This channel has the potential of explaining unusual features of events such as GW190521 \cite{Morton:2023wxg,LIGOScientific:2020iuh,Graham:2020gwr}.
It was also invoked to explain GW231123's unusually large masses \cite{Bartos:2025pkv,Delfavero:2025lup} and potential connections between flares and compact binary coalescences~\cite{Graham:2020gwr} (though evidence remains inconclusive; for example the kick direction of GW190412 is inconsistent with the AGN-CBC formation channel~\cite{Leong:2025qiw}).
Furthermore, AGNs could potentially be part of the answer to the so-called \enquote{final parsec problem} \cite{Milosavljevic:2002ht,Stone:2016wzz,Sayeb:2023vav}.
Recent statistical analyses of merging black hole populations have also shown hints in favor of this \ac{agn} formation channel \cite{Li:2025iux,Rowan:2022ehz,Rowan:2024lla}, with modest support for a link between high aligned binary spins appearing at high masses that persists even in data-driven analyses without strong physical assumptions, consistent with \ac{agn} formation~\cite{Rinaldi:2026nyb}.
Indeed, there is a growing body of evidence hinting at AGNs as a reasonable channel responsible for forming a fraction of the observed binary black holes.

\begin{figure}
    \centering
    \input{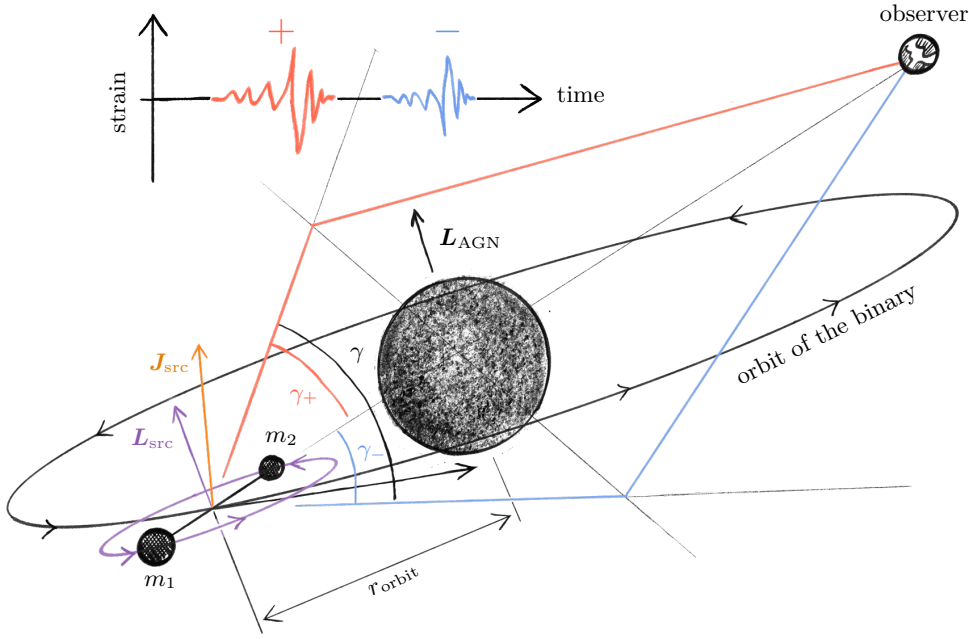} %
    \caption{
        Schematic diagram illustrating lensing in an AGN-CBC system.
        The supermassive black hole is here in an \ac{agn}, and the binary source is placed on a stable orbit at $\rorbit$ (\cref{fig:source_J_and_polarization_frames}), within the former's accretion disk.
        By symmetry (and in first approximation), the gravitational-wave emission would reach the observer via two paths (solid straight lines) and only two images would be observed, denoted by $+$ and $-$.
        Crucially, the total angle $\openangle \defdby \openanglep + \openanglem$ between their two emission lines (\cref{fig:lensing_plane_frame}) is \enquote{large} (\cref{eq:openanglepm_approx_on_LOS}).
        A schematic representation of the gravitational-wave strain the observer would observe is drawn on the top left (see also \cref{fig:lensed_waveforms}).
    }
    \label{fig:schematic_figure}
\end{figure}

If the merger occurs behind the central supermassive black hole of the \ac{agn}, relative to the observer, the gravitational waves are lensed and the gravitational-wave signal from the same source reaches the observer multiple times.
These repeated signals are the so-called \enquote{images} of the source.
Such a situation is illustrated in \cref{fig:schematic_figure}.
A recent work \cite{Leong:2024nnx} showed that the lensing probability of such an event is roughly inversely proportional to the radial distance of the \ac{cbc} from the supermassive black hole, which is expected to be much higher than the self-lensing by other potential objects, such as self-lensing by globular clusters (see the comprehensive study by~\cite{Ubach:2025dob}).
Therefore, if a sizable fraction of all binary mergers were to be part of an AGN-CBC, future observations are bound to see binary mergers lensed.
If not, this fraction can be strongly constrained.
It leaves the practical problem of determining whether a given gravitational-wave signal is lensed by a nearby supermassive black hole or not.
This work provides new tools to answer this question.

Typically, most lensed gravitational-wave systems have a small opening angle $\openangle$ (\cref{fig:schematic_figure}), which is why the two gravitational-wave signals essentially look the same~\cite{Hou:2019wdg}, but it was already recognized early on that the effect of Doppler shift is exacerbated near supermassive black holes~\cite{Itoh:2009iy,Yu:2021dqx,DOrazio:2019fbq,Toubiana:2020drf,Sberna:2022qbn,Kuntz:2022juv,Zhang:2023cdh,Samsing:2024xlo,Santos:2025ass,Postiglione:2025iys,Korzynski:2026iga}.
The effect was recently shown to be observable with LIGO-Virgo-KAGRA binary neutron star mergers~\cite{Zwick:2025nbg}.
Related work has also shown that, intriguingly, when a binary briefly passes behind a nearby massive black hole the gravitational-wave signal can temporarily brighten\,---much as a star brightens when it slides behind a lens in photometric microlensing---\,and that such a bump can constrain the orbital distance and the lens mass~\cite{Ubach:2025oxr}.
In addition to this, if the opening angle $\openangle$ (\cref{fig:schematic_figure}) is large, the effective angle of the binary's orbital plane is manifestly different for each image, which is a unique feature of AGN-CBC lensing~\cite{Korzynski:2026iga}.

\bigskip

This study considers what impact lensing would bear on the gravitational-wave signal; it connects compact binary coalescence's angular momentum (optionally) with the supermassive black hole's angular momentum.
The formalism is self-consistent with existing compact binary coalescence waveforms such that cutting-edge waveforms can be used, and it defines clear criteria to differentiate events lensed in an AGN-CBC from those lensed by other systems.
If such events were detected, here we show that the Doppler lensing effect and multi-view lensing encode unique information about the system, with up to four new unique constraints against one (two) new degrees of freedom when considering the binary's total/orbital angular momentum to be aligned with the supermassive black hole's due to accretion-disk-induced alignment of the binary's orbital properties (when considering the two to be independent).
The information includes the orbital distance to the supermassive black hole, which can constrain migration trap locations.
In addition, this stereoscopic view potentially offers a unique chance to peek into the higher-order modes of the \ac{cbc} like virtually no other known system can.
Indeed, higher-order modes\,---the subdominant modes of the spherical-harmonic decomposition of the \ac{gw} emission beyond the dominant quadrupole---\,are degenerate when viewed only along a single line of sight.
This degeneracy can be broken if one combines different points of view on the source.

\Cref{sec:system_setup} introduces the working assumptions and defines four distinct frames with respect to which all subsequent quantities are derived.
Once these are defined, the main results of this work are straightforward to obtain.
This is done in the subsequent two sections.
First, \cref{sec:geometrical_transformation_rules_of_angles} exhibits how the apparent parameters of each image differ, with respect to the orientation parameters one would infer in the absence of gravitational lensing.
Then, \cref{sec:Doppler_shift} computes the impact of the Doppler shift stemming from the same apparent orientations.
Finally, \cref{sec:discussion} discusses the results and concludes.

\section{System setup}
\label{sec:system_setup}

We consider the gravitational lensing of a gravitational-wave binary source by a nearby supermassive black hole.
The supermassive black hole can be that of an \ac{agn} and the source can be assumed to be embedded in its accretion disk, in which case, a number of simplifying assumptions are applicable.
However, we do neglect the mass of the accretion disk in our description of gravitational lensing.
The total mass $\Mbin$ of the binary is assumed to be much smaller than that $\MSMBH$ of the lensing black hole.
Indeed, while a supermassive black hole would typically have a mass of $\gtrapprox \qty{1e6}{\solarmass}$, the masses in binary mergers that current gravitational-wave observatories detect range from a few solar masses to \qty{\approx 1e2}{\solarmass}.
Similarly, we assume a hierarchy of distances to allow modelling the supermassive black hole by a point-like mass lens for gravitational waves to arrive in a near-planar regime.
The distance $d$ between the two merging bodies of the binary is assumed much smaller than the distance $\rorbit$ between their center of mass and that of the supermassive black hole: \ie. $d \ll \rorbit$, as is often done, at leading order, in b-EMRI systems.
The radius $\rorbit$ is itself much smaller than the distance between the observer and the lens (the supermassive black hole).
Previous works \cite{Leong:2024nnx,Bellovary:2015ifg,Graham:2020gwr,Peng:2021vzr,Secunda:2018kar,Thompson:2005mf} indicate that the distance $\rorbit$ matches that of migration traps, and it ranges from $\approx 10^1 \RS$ to $\approx 10^3 \RS$ (or, at most, $\rorbit \lessapprox 10^5 \, \RS$).
Here, $\RS$ designates the Schwarzschild radius of the central supermassive black hole, \ie.
\begin{equation}
    \RS = \frac{2 \, G \, \MSMBH}{c^{2}}\,,
    \label{eq:Schwarzschild_radius_definition}
\end{equation}
where $G$ is the gravitational constant, $c$ is the speed of light, and $\MSMBH$ is the mass of the supermassive black hole.
At leading order, the binary is put on a circular orbit of radius $\rorbit$ around the lens, and its speed $\vorbit$ reads
\begin{equation}
    \vorbit^2 = \frac{G \, \MSMBH}{\rorbit} = c^2 \frac{1}{2} \frac{\RS}{\rorbit} \,.
    \label{eq:vorbit_definition}
\end{equation}
Given the range of values for the mass $\MSMBH$ (\qtyrange{1e6}{1e8}{\solarmass}) and radius $\rorbit$ (\qtyrange{1e2}{1e4}{\RSunit}), the binary orbiting the supermassive black hole travels at speeds that are high enough to impart a measurable Doppler shift~\cite{Zhang:2024ibf}.
We shall thus incorporate it into our analysis; it is not to be confused with other effects, such as frame-dragging.

\subsection{Orbital, total angular momentum and polarization frames}
\label{sec:orbital_total_angular_momentum_and_polarization_frames}

\begin{figure}
    \centering
    \input{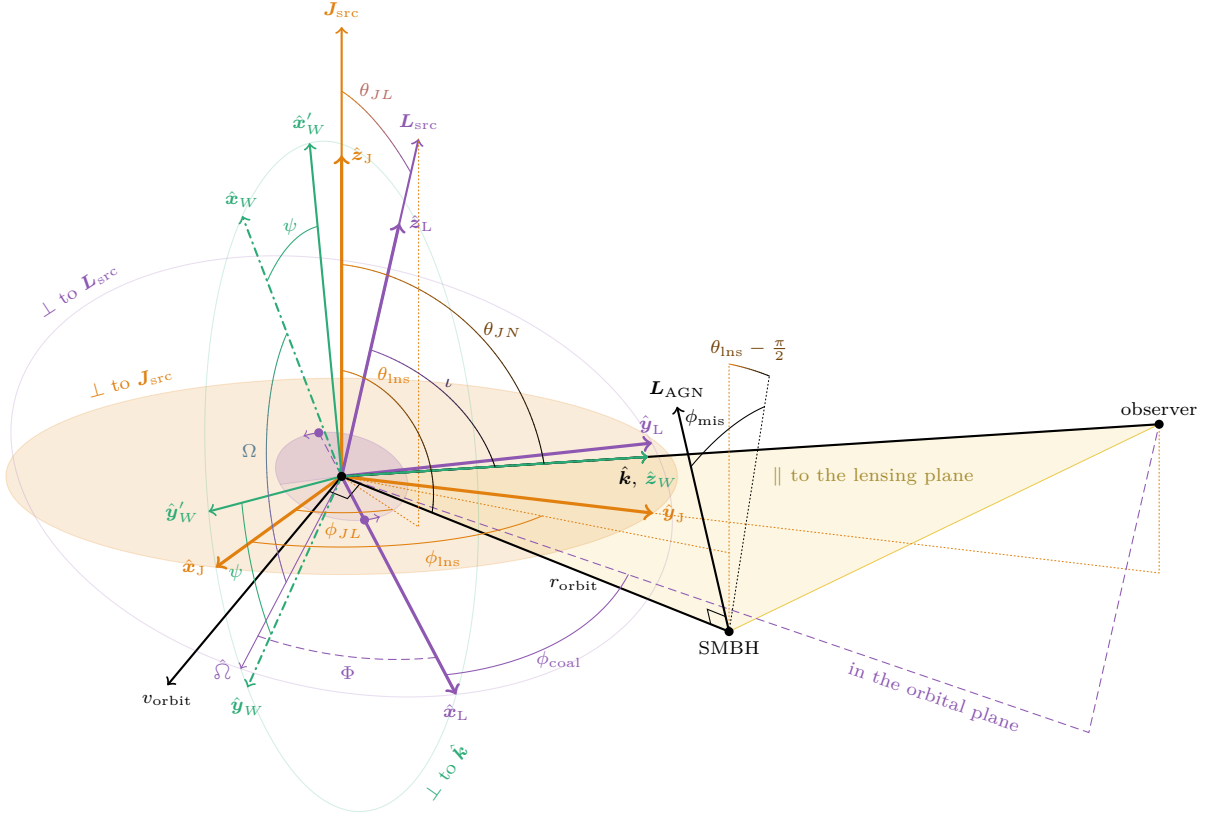} %
    \caption{
        Illustration of (1) the source frame $(\xhatL, \yhatL, \zhatL)$, (2) the total angular momentum frame $(\xhatJ, \yhatJ, \zhatJ)$ and (3) the polarization frame $(\uvec{x}_\text{W}, \uvec{y}_\text{W}, \uvec{z}_\text{W})$, along with (4) the definition of some angles.
        (1)
        The unit vector $\zhatL$ is defined to be along the orbital angular momentum $\Lvec$ of the binary, and the $\xhatL$ vector is defined such that it is along the axis defined by the two objects of the binary and pointing towards the secondary (lighter) of the two masses.
        (2)
        Similarly, $\zhatJ$ is along the total angular momentum $\Jvec$ of the binary, but $\xhatJ$ is then along the line of ascending nodes of the plane perpendicular to $\Jvec$ (or, equivalently, perpendicular to both $\Jvec$ and $\Nhat$, along $\Nhat \wedge \Jvec$).
        (3)
        To obtain the polarization frame, first define $(\uvec{x}'_\text{W}, \uvec{y}'_\text{W}, \uvec{z}'_\text{W})$, where $\uvec{z}'_\text{W} \defdby \Nhat$ and $\uvec{x}'_\text{W} \defdby \xhatJ$.
        The polarization frame $(\uvec{x}_\text{W}, \uvec{y}_\text{W}, \uvec{z}_\text{W})$ is then obtained by rotating this (primed) frame by an angle $-\psi$, around $\Nhat$.
        More details are given in \cref{sec:orbital_total_angular_momentum_and_polarization_frames} or \citet{Isi:2022mbx}.
        (4)
        The \enquote{inclination} $\incl$ is then the angle $\thetaLN$ between the observer position $\robsvec$ and the orbital angular momentum $\Lvec$.
        Similarly, one can define the angle $\thetaJN$ between $\Jvec$ and $\robsvec$.
        The difference between $\Lhat$ and $\Jhat$ is parametrized with the angles $\thetaJL$ and $\phiJL$ (\cref{app:angle_theta_JL}).
        The orbital phase is given by the angle between the line of ascending nodes in the orbital plane defined by $\Lvec$ and $\xhatL$ (with an $\Omega\defdby\frac{\pi}{2}$ shift).
        Its value at the coalescence time $\tcoal$ sets the phase of coalescence, \ie. $\phicoal \defdby \phase(t=\tcoal) = \frac{\pi}{2} -\phiorb(t=\tcoal)$.
        Finally, the lensing supermassive black hole is located in the total angular momentum frame via the angles $\thetalns$ and $\philns$ (and the distance $\rorbit$).
        Similarly, the rotation axis of the binary is along $\LAGNvec$, whose direction is given by the angles $\thetamis = \frac{\pi}{2}$ and $\phimis$.
        Some explicit expressions are given in \cref{app:explicit_definition_of_the_basis_vectors}, and \cref{app:highglighting_the_angles_that_change,app:highglighting_the_new_parameters} highlight particular aspects of the illustration.
    }
    \label{fig:source_J_and_polarization_frames}
\end{figure}

In the computations that follow, we will juggle with different coordinate frames and their basis vectors.
These definitions are illustrated in \cref{fig:source_J_and_polarization_frames,fig:lensing_plane_frame}, and some explicit expressions are given in \cref{app:explicit_definition_of_the_basis_vectors}.
Whenever possible, we follow the conventions adopted by the \ac{lvk} collaboration, and match the definitions given in \citet{Isi:2022mbx}.

Firstly, we introduce the $\Lvec$-based orbital frame or source frame $(\xhatL,\yhatL,\zhatL)$.
By definition, $\zhatL \parallel \Lvec$ and $\xhatL$ points from the primary object (heavier) towards the secondary one (lighter).
It is centered on the center of mass of the binary and its definition relies only on the binary's own intrinsic properties, as \cref{fig:source_J_and_polarization_frames} shows.

Secondly, a similar frame $(\xhatJ, \yhatJ, \zhatJ)$ can be constructed by using the total angular momentum $\Jvec$ in lieu of the orbital angular momentum $\Lvec$, but where $\xhatJ$ and $\yhatJ$ are obtained relative to the observer position $\robsvec$.
So, unlike the source frame, this total angular momentum frame depends on both the source and observer's locations.

Thirdly, in order to describe the polarization angle $\psi$, we make use of yet another frame in our computations: the polarization frame $(\uvec{x}_W, \uvec{y}_W, \uvec{z}_W)$, also described in \cref{fig:source_J_and_polarization_frames}~\cite{Isi:2022mbx}.
This frame orientates the orbital plane of the binary with respect to the celestial coordinates used to localize the event in the Earth's sky.

\bigskip

To talk about the Doppler shift, we do not only need the orbital speed amplitude (\cref{eq:vorbit_definition}), but also its direction.
Provided that the binary black hole is on a stable orbit around the supermassive black hole, within its accretion disk, itself rotating along the spin of the \ac{agn}, some statements can be made.
The radial axis between the gravitational-wave source and the lensing supermassive black hole, the angular momentum of the accretion disk $\LAGNvec$, and the velocity of the binary $\vorbitvec$ are all perpendicular to each other.
We can thus describe the orientation of the binary with respect to the supermassive black hole with the misalignment angle $\phimis$.
If $\thetalns=\frac{\pi}{2}$, then $\phimis$ is the angle between the binary's total angular momentum $\Jvec$ and the accretion disk's angular momentum $\LAGNvec$.
Then, only if both $\phimis = 0$ and $\thetalns = \frac{\pi}{2}$, the two angular momenta are aligned.
The angular momentum of the \ac{agn} and the spin of the supermassive black hole are usually aligned~\cite{Yang:2019cbr,Li:2025iux}.
Notice that one is free to use different frames, such as the source frame, to measure these angles instead.

\subsection{Lensing plane and lensing frame}
\label{sec:lensing_plane_and_lensing_frame}

We now turn our attention to the lensing plane, \ie. the plane within which the gravitational-wave signals that reach the observer\,---in the geometric optics limit---\,propagate, and an associated frame.
By symmetry, it is the plane that includes the observer, the lens and the source.
The angles and distances usually defined in gravitational lensing, within the lensing plane, are made explicit in \cref{fig:lensing_plane_frame}.
Notice there that, in general, the angular distance $\Ds$ is \emph{not} equal to $\Dls + \Dl$.
However, in the present configuration, since the source is placed very close to the lens, \ie. $\Dls \ll \Ds \approx \Dl$, this sum is a valid approximation.
Similarly, this scale of distances also brings in other simplifications in the expression of lensing parameters (\eg., \cref{app:asymptotic_expressions_for_an_infinitely_far_away_observer}).
To work with the lensing plane, we introduce a fourth and last frame $(\xhatlns, \yhatlns, \zhatlns)$, which we call the \enquote{lensing frame}, and which is defined in \cref{fig:lensing_plane_frame}.

\begin{figure}
    \centering
    \tikzsetnextfilename{lensing_plane_frame}
\begin{tikzpicture}[thick,line cap=round,scale=1.3]
    \newcommand{\lenscolor}{RoyalBlue}
    \newcommand{\sourcecolor}{ForestGreen}

    \def\Dl{4.8}
    \def\Dls{4.4}
    \def\lenseta{0.65}
    \def\lensxi{1.25}

    \pgfmathsetmacro\Ds{ (\Dl+\Dls) }
    \pgfmathsetmacro\imageyup{\lensxi/\Dl*\Ds}
    \pgfmathsetmacro\anglethetaup{atan(\lensxi/\Dl)}
    \pgfmathsetmacro\anglebeta{atan(\lenseta/\Ds)}
    \pgfmathsetmacro\anglethetaE{sqrt(\anglethetaup^2-\anglebeta*\anglethetaup)}
    \pgfmathsetmacro\lensRE{tan(\anglethetaE)*\Dl}
    \pgfmathsetmacro\lenssize{tan(\anglethetaE)*\Dl}
    \pgfmathsetmacro\anglethetadn{-\anglethetaup+2*sqrt(\anglebeta^2/4+\anglethetaE^2)}
    \pgfmathsetmacro\lensxidn{tan(\anglethetadn)*\Dl}
    \pgfmathsetmacro\imageydn{\lensxidn/\Dl*\Ds}
    \pgfmathsetmacro\orbitalradius{sqrt(\lenseta*\lenseta+\Dls*\Dls)}
    \pgfmathsetmacro\anglealphaupminusthetaup{atan( ( tan(\anglethetaup) - tan(\anglebeta) )*\Ds/\Dls - tan(\anglethetaup) )}
    \pgfmathsetmacro\anglealphadnminusthetadn{atan( ( tan(\anglethetadn) + tan(\anglebeta) )*\Ds/\Dls - tan(\anglethetadn) )}

    \pgfmathsetmacro\observerup{ \Ds * (\lensxi-\lenseta) / \Dls + \lenseta}
    \pgfmathsetmacro\observerdn{ \Ds * (\lensxidn+\lenseta) / \Dls - \lenseta}

    \coordinate (src) at (0,\lenseta);
    \coordinate (lns) at (\Dls,0);
    \coordinate (obs) at (\Ds,0);

    \coordinate (imgup) at (0,\imageyup);
    \coordinate (imgdn) at (0,-\imageydn);
    \coordinate (obsup) at (\Ds,\observerup);
    \coordinate (obsdn) at (\Ds,-\observerdn);

    \draw[LnsPlaneCol, very thin] (src) -- (obs) -- (lns) -- cycle;
    \fill[LnsPlaneCol, opacity=0.15] (src) -- (obs) -- (lns) -- cycle;

    \node[\sourcecolor,left] at (\Ds-4,0.15) {\scriptsize $\beta$};
    \draw[\sourcecolor,thin] (obs) ++(0:-4) arc (180:180-\anglebeta:4);

    \node[\imagecolor,left] at (\Ds-3.5,0.55) {\scriptsize $\alpha_+$};
    \draw[\imagecolor,thin] (obs) ++(180-\anglebeta:3.6) arc (180-\anglebeta:180-\anglethetaup:3.6);

    \node[\imagecolordn,left] at (\Ds-4.3,-0.55) {\scriptsize $\alpha_-$};
    \draw[\imagecolordn,thin] (obs) ++(180-\anglebeta:4.4) arc (180-\anglebeta:180+\anglethetadn:4.4);

    \node[\imagecolor,left] at (\Ds-3,0.52) {\scriptsize $\theta_+$};
    \draw[\imagecolor,thin] (obs) ++(0:-3) arc (180:180-\anglethetaup:3);

    \node[\imagecolordn,left] at (\Ds-3.5,-0.45) {\scriptsize $\theta_-$};
    \draw[\imagecolordn,thin] (obs) ++(0:-3.5) arc (180:180+\anglethetadn:3.5);

    \node[Black,opacity=0.4,left] at (\Ds-2.5,0.42) {\scriptsize $\thetaE$};
    \draw[Black,opacity=0.4,thin] (obs) ++(0:-2.5) arc (180:180-\anglethetaE:2.5);

    \node[Black,opacity=0.4,left] at (\Ds-2.7,-0.35) {\scriptsize $\thetaE$};
    \draw[Black,opacity=0.4,,thin] (obs) ++(0:-2.7) arc (180:180+\anglethetaE:2.7);

    \node[\imagecolor,left] at (\Dls-1,\lensxi+0.05) {\scriptsize $\hat{\alpha}_+$};
    \draw[\imagecolor,thin] (\Dls,\lensxi) ++(0:-1) arc (180:180-\anglethetaup:1);
    \draw[\imagecolor,thin] (\Dls,\lensxi) ++(0:-1) arc (180:180+\anglealphaupminusthetaup:1);

    \node[\imagecolordn,left] at (\Dls-1,-\lensxidn+0.05) {\scriptsize $\hat{\alpha}_-$};
    \draw[\imagecolordn,thin] (\Dls,-\lensxidn) ++(0:-1) arc (180:180+\anglethetadn:1);
    \draw[\imagecolordn,thin] (\Dls,-\lensxidn) ++(0:-1) arc (180:180-\anglealphadnminusthetadn:1);

    \node[\sourcecolor,right] at (0+0.4,\lenseta+0.25) {\scriptsize $\openangle$};
    \draw[\sourcecolor,thin] (src) ++(0:0.6) arc (0:\anglealphaupminusthetaup:0.6);
    \draw[\sourcecolor,thin] (src) ++(0:0.6) arc (0:-\anglealphadnminusthetadn:0.6);
    \node[\sourcecolor,right] at (0+0.55,\lenseta-0.45) {\scriptsize $\openanglem$};
    \draw[\sourcecolor,thin] (src) ++(-\anglebeta:0.75) arc (-\anglebeta:-\anglealphadnminusthetadn:0.75);
    \node[\sourcecolor,right] at (0+1.6,\lenseta+0.05) {\scriptsize $\openanglep$};
    \draw[\sourcecolor,thin] (src) ++(-\anglebeta:1.6) arc (-\anglebeta:\anglealphaupminusthetaup:1.6);

    \draw[\sourcecolor, dotted] (\Dl+\Dls, 0) -- (src);

    \draw[\imagecolor, very thick] (src) -- (\Dls,\lensxi);
    \draw[\imagecolor, very thick] (\Dls,\lensxi) -- (obs);
    \draw[Black, opacity=0.4, thin] (\Dls-0.15,{tan(\anglethetaE)*(\Dl+0.15)}) -- (obs);
    \draw[Black, opacity=0.4, thin] (\Dls-0.15,{-tan(\anglethetaE)*(\Dl+0.15)}) -- (obs);
    \draw[\imagecolor, very thick, densely dashed] (imgup) -- (\Dls,\lensxi);
    \draw[\imagecolor, thick, dotted] (\Dls,\lensxi) -- (obsup);

    \draw[\imagecolordn, very thick] (src) -- (\Dls, -\lensxidn);
    \draw[\imagecolordn, very thick] (\Dls,-\lensxidn) -- (obs);
    \draw[\imagecolordn, very thick, densely dashed] (imgdn) -- (\Dls,-\lensxidn);

        \draw[\imagecolordn, thick, dotted] (\Dls,-\lensxidn) -- (obsdn);

    \fill[\imagecolor] (imgup) circle (1.5pt);
    \fill[\imagecolordn] (imgdn) circle (1.5pt);
    \fill[\imagecolor] (obsup) circle (1.5pt);
    \fill[\sourcecolor] (src) circle (1.5pt);
    \fill[Navy] (obs) circle (1.5pt);
    \fill[Black] (lns) circle (1.5pt);

    \node[below left] at (src) {\scriptsize source};
    \node[above left] at (imgup) {\scriptsize image $+$};
    \node[left] at (imgdn) {\scriptsize image $-$};
    \node[above left] at (obsup) {\scriptsize observer $+$};
    \node[below] at (obs) {\scriptsize observer};
    \node[below] at (lns) {\scriptsize lens};

    \draw[Black,opacity=0.4,densely dotted,thin] (\Dls-0.5,\lensRE) -- (\Dls+0.25,\lensRE);
    \draw[Black,opacity=0.4,densely dotted,thin] (\Dls-0.5,-\lensRE) -- (\Dls+0.25,-\lensRE);

    \draw[Black, opacity=0.4, thin, <->] (\Dls-0.2,0) -- (\Dls-0.2,\lensRE);
    \draw[Black, opacity=0.4, thin, <->] (\Dls-0.2,0) -- (\Dls-0.2,-\lensRE);

    \node[Black, opacity=0.4, left] at (\Dls-0.2, \lensRE/2+0.2) {\scriptsize $\RE$};
    \node[Black, opacity=0.4, left] at (\Dls-0.2, -\lensRE/2+0.1) {\scriptsize $\RE$};

    \draw[<->] (0,-1.3) -- (\Dls,-1.3);
    \draw[<->] (\Dls,-1.3) -- (\Ds,-1.3);
    \draw[<->] (0,-1.45) -- (\Ds,-1.45);

    \draw[Black,opacity=0.4,densely dotted,thin] (0,-\imageydn-0.25) -- (0,\imageyup+0.25);
    \draw[Black,opacity=0.4,densely dotted,thin] (\Ds,{-\observerdn-0.25}) -- (\Ds,{\observerup+0.25});
    \draw[Black,opacity=0.4,densely dotted,thin] (\Dls,{-\imageydn-0.25}) -- (\Dls,\lensxi+0.25);
    \draw[Black,opacity=0.4,densely dotted,thin] (0-1.9,0) -- (\Ds+0.5,0);

    \node[below, align=center] at (0,{-\imageydn-0.25}) {\scriptsize source \\ \scriptsize plane};
    \node[below, align=center] at (\Dls,{-\imageydn-0.25}) {\scriptsize lens \\ \scriptsize plane};

        \fill[\imagecolordn] (\Ds,-\observerdn) circle (1.5pt);
        \node[left] at (\Ds,-\observerdn) {\scriptsize observer $-$};

    \node[above left] at (\Dls/2,-1.3) {\scriptsize $D_{\ell s}$};
    \node[above] at (\Dls+\Dl/2+0.5,-1.3) {\scriptsize $D_\ell$};
    \node[below] at (\Ds/2,-1.45) {\scriptsize $D_{s}$};

    \draw[<->] (-1.2,0) -- (-1.2,\lenseta);
    \draw[<->] (-1.0,0) -- (-1.0,\imageyup);
    \draw[Black,opacity=0.4,densely dotted,thin] (-1.5,\lenseta) -- (0.5,\lenseta);
    \draw[Black,opacity=0.4,densely dotted,thin] (-1.5,\imageyup) -- (0.5,\imageyup);

    \node[left] at (-1.0,\lensxi/2+\imageyup/2) {\scriptsize $\xi_+ \equiv x_+ \cdot \RE$};
    \node[left] at (-1.2,\lenseta/2) {\scriptsize $\eta \equiv y \cdot\RE$};

    \draw[thin, densely dotted] (0, \lenseta) -- (\Dls, 0);
    \node[right] at ({\Dls/2 + 1.1}, {\lenseta/2 - 0.1}) {\scriptsize $\rorbit$};

    \node[right] at ({cos(\anglebeta)}, {\lenseta - sin(\anglebeta) + 0.11}) {\scriptsize $\xhatlns$};
    \draw[->] (0, \lenseta) -- ({cos(\anglebeta)}, {\lenseta - sin(\anglebeta)});
    \node[above] at ({sin(\anglebeta)}, {\lenseta + cos(\anglebeta)}) {\scriptsize $\yhatlns$};
    \draw[->] (0, \lenseta) -- ({sin(\anglebeta)}, {\lenseta + cos(\anglebeta)});

\end{tikzpicture}
    \caption{
        The above diagram situates the quantities used in the text, in the lensing plane.
        Notice that, unlike how it is often assumed when studying gravitational lensing, not all angles can be assumed small.
        The distances $\Dls$, $\Dl$ and $\Ds$ denote the angular distances between the lens and the source, the lens and the observer, and the source and the observer, respectively.
        The basis unit vector $\xhatlns$ is aligned from the source towards the observer, and $\yhatlns$ is perpendicular, within the lensing plane, and points \enquote{upwards} from the lens.
        The last unit vector $\zhatlns$ completes the triad.
        The opening angle $\openangle = \openanglep+\openanglem$ is represented in green on the left.
        In the geometric optics limit, the two trajectories of the gravitational waves that reach the observer are traced in solid red (above the lens) and solid blue (below the lens).
        See also \cref{app:basis_vectors_of_the_lensing_frame} for some explicit definitions.
    }
    \label{fig:lensing_plane_frame}
\end{figure}

The resolution of the lens equation for a point mass is a typical textbook example \cite{Weinberg:2008zzc,congdonPrinciplesGravitationalLensing2018,poissonGravityNewtonianPostNewtonian2014,willTheoryExperimentGravitational2018}.
The lens equation $\alpha_\pm + \beta = \theta_\pm$ can here be exactly solved, and it yields
\begin{equation}
    \theta_\pm = \frac{\beta}{2} \pm \sqrt{\frac{\beta^2}{4}+\thetaE^2} \,,
    \label{eq:theta_pm_solutions}
\end{equation}
where the Einstein angle $\thetaE$ is defined by \cite{Weinberg:2008zzc}
\begin{equation}
    \thetaE^2 \defdby \frac{4G \MSMBH}{c^2} \frac{\Dls}{\Dl \Ds}
    = 2 \RS \frac{\Dls}{\Dl \Ds}\; .
    \label{eq:Einstein_angle_definition}
\end{equation}
An Einstein radius $\RE$ can be associated with this angle.
Physically speaking, it corresponds to the radius of the Einstein ring around the lens subtended by the Einstein angle $\thetaE$ when the source, the lens and the observer are all perfectly aligned.
Explicitly, it thus reads
\begin{equation}
    \RE \defdby\thetaE\Dl\,.
    \label{eq:Einstein_radius_definition}
\end{equation}

In parallel, after being lensed, the two signals' amplitudes are magnified by a factor $\mu_+$ and $\mu_-$, respectively.
These can be found to be \cite{Weinberg:2008zzc,congdonPrinciplesGravitationalLensing2018}
\begin{equation}
    \mu_\pm
    =
    \left|
        \frac{\theta_\pm}{\beta} \frac{\dd \theta_\pm}{\dd \beta}
    \right|
    =
    \left|
        1 - \frac{\thetaE^4}{\theta_\pm^4}
    \right|^{-1} \,.
    \label{eq:magnification_factors}
\end{equation}

To summarize, we defined the following 4 frames (see \cref{fig:source_J_and_polarization_frames,fig:lensing_plane_frame}, as well as \citet{Isi:2022mbx}), which were all chosen to be centered on the gravitational-wave source binary.
\begin{enumerate}
    \item The orbital $\Lvec$-frame is intrinsic to the binary source, and only depends on $\Lvec$ (and the mass locations).
    \item The $\Jvec$-frame ties together the total angular momentum $\Jvec$, the source position and the observer's direction $\Nhat$.
    \item The polarization frame relates the orientation of the celestial coordinate frame, used to localize the source in the sky, with the orientation of the orbital plane.
    \item The lensing frame describes the gravitational-wave propagation paths and helps determine the image or virtual observer directions.
\end{enumerate}

\section{Geometrical transformation rules of angles}
\label{sec:geometrical_transformation_rules_of_angles}

In the absence of any lensing, a (quasi-circular) \ac{cbc} is typically characterized by $15$ parameters.
The properties intrinsic to the source itself are given by 8 parameters\footnote{We follow the parametrization employed in the Python package Bilby \cite{Ashton:2018jfp,bilby_doi}.}, which include $2$ masses $m_1$ and $m_2$, $2$ spin amplitudes, $2$ spin tilts, and a phase angle between the spins.
By specifying an eighth parameter, such as the phase angle $\phiJL$ between the total angular momentum $\Jvec$ and the orbital angular momentum $\Lvec$, all other quantities intrinsic to the sources are now defined.
The detector introduces a further $4$ extrinsic parameters: the inclination angle $\incl$, the phase of coalescence $\phicoal$, the coalescence time, and the luminosity distance $\dL$.
Sometimes, instead of the inclination $\incl$, it is the angle $\thetaJN$ between the observer's direction $\Nhat$ and the total angular momentum $\Jvec$ that is used.
Two extra parameters determine the sky localization: the right ascension $\ra$ and the declination $\dec$.
The last parameter we have not mentioned until now is the polarization angle $\psi$, which is defined relative to the observer and the source \cite{Isi:2022mbx}.
In total, one now counts $15$ parameters.

In case of lensing by a supermassive black hole, \cref{fig:source_J_and_polarization_frames} introduced a number of new parameters to include a lens in the description of the overall system.
We must include the mass $\MSMBH$ of the lens, the distance $\rorbit$ between the lens and the source, and the two angles $\philns$ and $\thetalns$ to situate the lens with respect to the source (\cref{fig:source_J_and_polarization_frames}).
As we assume the source to be on a stable circular orbit within the accretion disk of the \ac{agn}, the angle $\phimis$ alone is enough to complete the description of the system.
This angle $\phimis$ thus tells us how the source is moving around the lens, and will be reflected in the Doppler shift imparted to the lensed signals.
With these $5$ new parameters (\cref{fig:new_parameters_only}), we now have $20$ parameters to take care of.

However, in principle, some simplifying assumptions can reasonably be made to reduce their number.
If we assume that either the orbital angular momentum $\Lvec$ or the total one $\Jvec$ is aligned with that of the accretion disk, $\LAGNvec$, within which the binary is embedded, we may remove two degrees of freedom.
For example, aligning the total angular momentum $\Jvec$ with $\LAGNvec$ means that $\thetalns = \frac{\pi}{2}$ and $\phimis = 0$.
If that is the case, lensing of a gravitational-wave source in an \ac{agn} adds $3$ parameters on top of the $15$ gravitational-wave parameters, or only $1$ more than if the source was lensed by a point-mass lens\footnote{For a point-mass lens, the impact of lensing is fully determined by the dimensionless source position $y \defdby \eta \cdot \RE^{-1}$ (\cref{eq:Einstein_radius_definition,fig:lensing_plane_frame}) and the redshifted lens mass.}.
This assumption can be motivated by the formation dynamics expected for such a binary in an \ac{agn}, where misaligned angular momenta can be dragged into alignment \cite{Yang:2019cbr,Li:2025iux}.
One could go further and argue that the spins of the two masses in the binary should also be aligned.
If so, that removes a further $4$ degrees of freedom: the two spin tilts are then null, and their phase difference becomes irrelevant, and so does $\phiJL$ since aligned spins means $\Jvec \parallel \Lvec$.
Therefore, in the simplest case, the system can thus be considered characterized by only 14 parameters.
While the calculations described in this work have been carried out without these assumptions, these simplifying assumptions will sometimes be invoked to provide shorter explicit expressions.

In practice, lensing will bias some parameters, while leaving others unchanged.
First, as \cref{fig:lensing_plane_frame} makes clear, the observer sees two distinct images of the source arriving at two different times.
In principle, these two images originate from two different regions of the sky.
However, the resolution of current (and future) gravitational-wave observatories is not precise enough to resolve these two locations distinctly \cite{Hannuksela:2020xor,ET:2019dnz,Evans:2023euw}.
Therefore, practically speaking, the two events would share the same apparent sky location.
In the case of lensing, the luminosity distances inferred from each gravitational-wave signal must also be biased by the magnification factors $\mu_+$ and $\mu_-$ (\cref{eq:magnification_factors}), and the second image $-$ will have its phase shifted by a constant amount; this is the Morse phase shift of so-called Type II (or secondary) images~\cite{Schneider:1992bmb}.

\begin{figure}
    \centering
    \input{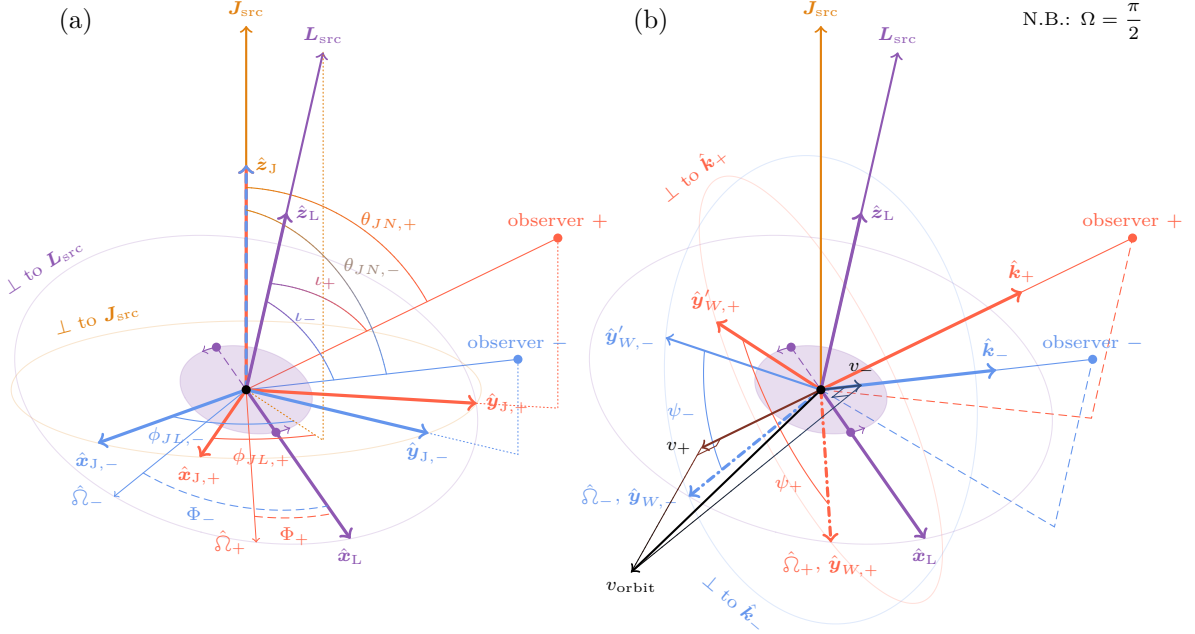}
    \caption{
        Illustration of how the parameters transform when the observer's position changes between two distinct perspectives: an \enquote{observer $+$} or an \enquote{observer $-$}.
        For a complete definition of all parameters, refer to \cref{fig:source_J_and_polarization_frames}.
        For clarity, the illustration is split into two panels, left (a) and right (b), that share an identical configuration; for convenience, $\Omega $ was chosen to be $\frac{\pi}{2}$, following the \ac{lvk} Collaboration's convention.
        The left panel (a) exhibits how the $\Jvec$-frame changes with the observer's position (notice however that $\zhat = \uvec{z}_+ = \uvec{z}_-$).
        This modifies the $\phiJL$ azimuthal angle, the orbital and coalescence phases $\phiorb \equiv \frac{\pi}{2} - \phicoal$, as well as the inclinations $\thetaJN$ and $\incl$ with respect to the $\Jvec$ and $\Lvec$ momenta, respectively.
        The right panel (b) shows how the polarization angle changes\,---in lensing specifically---\,and how the orbital velocity $\vorbitvec$ gets projected along the two distinct lines of sight.
        See also \cref{app:explicit_definition_of_the_basis_vectors} for some explicit definitions.
    }
    \label{fig:changes_with_two_perspectives}
\end{figure}

In the lensing of a gravitational-wave source inside an \ac{agn}, we now argue that other parameters will vary.
To determine which ones get modified, first consider what changes between two distinct points of view on the same binary merger event; we denote with $+$ and $-$ the two points of view such that the $+$ corresponds to the primary image (\ie. the first to arrive) and the $-$ to the secondary one.
For example, the intrinsic masses involved have no reason to be modified by the viewing orientation.
\Cref{fig:changes_with_two_perspectives,fig:angle_definitions_in_isolation} illustrate which vectors depend on the observer's direction and how they change in case of lensing;
while the orbital frame and the lensing frame do not change---the former is intrinsic to the source and the latter is fixed by the source, lens, and observer positions alone.
Given the source frame definition (\cref{fig:source_J_and_polarization_frames}), it is straightforward to see that the inclination angle $\incl$ (or $\thetaJN$) and the phase of coalescence $\phicoal$ have to change.
Indeed, they both depend on the effective direction $\Nhatpm$ of the line of sight.
Notice, however, that, when one is maximally biased, the other is not biased at all.
It turns out that, adopting the same definition as \ac{lvk}, the polarization angle $\psi$ also changes \cite{Isi:2022mbx}.
Finally, in view of \cref{eq:vorbit_definition}, we also want to incorporate the possible Doppler shift that the orbital speed may bring in.
Effectively, this will bias the two masses of the binary by introducing a shift in the frequency of the gravitational-wave signal.
In what follows, we will exhibit the transformation rules to obtain the two lensed signal parameters, given the \enquote{unlensed} parameters (\ie. what is observed/measured without the lens in place).

In order, this section describes what changes must be applied to $\thetaJN$, the angle $\phiJL$, the phase of coalescence $\phicoal$ and the polarization angle $\psi$.
Whenever necessary, we will make use of the angle $\thetaJL$, \ie. the angle between $\Jvec$ and $\Lvec$ (\cref{app:angle_theta_JL}).
Strictly speaking, this angle $\thetaJL$ is not independent, but its determination requires taking into account the spin of the two masses, and since we are taking a geometric description, we here prefer to use it instead. %

The two effective points of view on the source (\cref{fig:lensing_plane_frame}) lie in the lensing plane and are directed, in terms of $\xhatlns$ and $\yhatlns$, towards
\begin{equation}
    \Nhatpm
        = R_{\zhatlns}(\openanglepm) \cdot \xhatlns
        = \cos(\openanglepm) \cdot \xhatlns
        \pm \sin(\openanglepm) \cdot \yhatlns \,,
    \label{eq:virtual_observer_directions_definition}
\end{equation}
where $R_{\zhatlns}(\openanglepm)$ denotes the rotation matrix around $\zhatlns$ by an angle $\openanglepm$, and where we defined
\begin{equation}
    \openanglepm \defdby \hat{\alpha}_\pm - \theta_\pm \pm \beta \,.
    \label{eq:openanglepm_definition}
\end{equation}
Notice that $\openanglep + \openanglem = \openangle$, the (total) opening angle, as indicated in \cref{fig:lensing_plane_frame}.
Furthermore, for a far-away observer $\Dls \ll \Ds \approx \Dl$, the angle $\openanglepm$ can be expressed independently of $\Dl$ (or $\Ds$), as given in \cref{app:gamma_pm_for_a_far_away_observer}.
Similar statements can be made for the corresponding magnifications $\mu_\pm$ and arrival time difference (\cref{app:asymptotic_expressions_for_an_infinitely_far_away_observer}).

We now make a short parenthesis to calculate a rough estimate for the opening angle $\openangle$.
For simplicity, let us assume that the source, the lens and the observer are almost perfectly aligned, \ie. $\beta \approx 0$.
Then, by symmetry, $\openanglep \approx \openanglem \approx \frac{\openangle}{2}$ and $\theta_+ \approx \theta_-$ (actually, if $\beta \equiv 0$, we do not have two images, but a ring).
Since $\theta_\pm$ is small, we can write
\begin{equation}
    \Dl \, \theta_\pm \approx \tan(\openanglepm) \Dls \,.
\end{equation}
Substituting $\theta_\pm$ by its solution (\cref{eq:theta_pm_solutions}), making $\thetaE$ explicit (\cref{eq:Einstein_angle_definition}), and using $\Ds \approx \Dl$, we obtain
\begin{equation}
    \tan(\openanglepm) \approx \sqrt{\frac{2 \RS}{\Dls}} \,.
    \label{eq:openanglepm_approx_on_LOS}
\end{equation}
If $\Dls$ ranges from $10^1\,\RS$ to $10^3\,\RS$, we find that $\openangle$ can go from about \qty{50}{\degree} to \qty{5}{\degree}, respectively.
Compared to usual lensing scenarios, where $\Dls$ is of similar magnitude to $\Dl$, these are relatively large angles, which motivates the present study.
This ends our parenthesis.

\subsection{Inclination \texorpdfstring{$\thetaJN$}{θJN} of the total angular momentum}

Using \cref{eq:virtual_observer_directions_definition}, the change in $\thetaJN$ can be compactly expressed by
\begin{equation}
    \cos(\thetaJNpm) = \zhatJ \cdot \Nhatpm \,.
    \label{eq:cos_thetaJNpm_definition}
\end{equation}
In explicit terms, the above reads
\begin{equation}
    \cos(\thetaJNpm) =
    \cos (\openanglepm) \cos (\theta_{JN})
    \pm
    \frac{\sin (\openanglepm) \sin (\thetaJN) \bigl(
        \sin (\philns) \cos (\thetaJN) \sin (\thetalns)-\sin (\thetaJN) \cos (\thetalns)
    \bigr)}{\sqrt{\denomInJFrame}} \,,
    \label{eq:cos_thetaJNpm_explicit}
\end{equation}
where $\sqrt{\denomInJFrame}$ is given by
\begin{equation}
    \sqrt{\denomInJFrame}
    \defdby
    \frac{\norm{(\robsvec-\rsrcvec)\wedge(\rsrcvec-\rlnsvec)}}{\norm{\robsvec}\cdot\norm{\rlnsvec}}
    =
    \norm{
    \begin{pmatrix}
        \cos(\thetaJN) \sin(\philns) \sin(\thetalns) - \cos(\philns) \sin(\thetaJN) \\
        -\cos(\philns) \cos(\thetaJN) \sin(\thetalns) \\
        \cos(\philns) \sin(\thetaJN) \sin(\thetalns)
    \end{pmatrix}
    } \,,
    \label{eq:denomInJFrame_definition}
\end{equation}
which comes up naturally through the definition of the lensing frame.
The choice of localization angles for the lens (\ie. $\philns$ and $\thetalns$) makes \cref{eq:cos_thetaJNpm_explicit} probably the most straightforward of all angle transformations.
If $\Jvec \parallel \LAGNvec$, then $\thetalns=\frac{\pi}{2}$, and at first order in $\openanglepm$, the expression reduces further down to
\begin{equation}
    \cos(\thetaJNpm) \biggr\rvert_{\Jvec \parallel \LAGNvec}
    =
    \cos (\theta_{JN})
    \pm
    \frac{\openanglepm \sin (\philns) \sin (\thetaJN) \cos (\thetaJN)}{\sqrt{\denomInJFrame}}
    +\bigOof{\openanglepm^2} \,.
    \label{eq:cos_thetaJNpm_approx_if_L_and_J_aligned}
\end{equation}

If instead of expanding the cosine (\cref{eq:cos_thetaJNpm_definition}), one were to invert the expression and directly expand the angle, the result would read
\begin{equation}
    \thetaJNpm
    =
    \thetaJN
    \pm
    \frac{\openanglepm (\sin (\thetaJN) \cos (\thetalns)-\sin (\philns) \cos (\thetaJN) \sin (\thetalns))}{\sqrt{\denomInJFrame}}+\bigOof{\openanglepm^2} \,,
    \label{eq:thetaJNpm_approx}
\end{equation}
and, if the angular momentum $\Jvec$ of the binary aligns with that of the accretion disk $\LAGNvec$, it becomes
\begin{equation}
    \thetaJNpm \biggr\rvert_{\Jvec \parallel \LAGNvec}
    =
    \thetaJN
    \mp \frac{\openanglepm  \sin (\philns) \cos (\thetaJN)}{\sqrt{\denomInJFrame}}+\bigOof{\openanglepm^2} \,. %
    \label{eq:thetaJNpm_approx_if_L_and_J_aligned}
\end{equation}

As mentioned previously, the inclination $\thetaLN \equiv \iota$ can also, in principle, be used in lieu of $\thetaJN$.
Similarly to the change in $\thetaJN$ (\cref{eq:cos_thetaJNpm_definition}), the change in inclination $\thetaLN \equiv \incl$ can be written only in terms of $\incl$, $\philns$ and $\thetalns$, as
\begin{equation}
    \cos(\incl_\pm) = \zhatL \cdot \Nhatpm \,.
    \label{eq:cos_incl_definition}
\end{equation}
Assuming that $\thetalns = \frac{\pi}{2}$, at first order in small angles, it can easily be shown to naturally reduce to \cref{eq:cos_thetaJNpm_approx_if_L_and_J_aligned}.

\subsection{Angle \texorpdfstring{$\phiJL$}{ϕJL}}

If the total angular momentum and the orbital angular momentum are misaligned, \ie. $\thetaJL \neq 0$, then their relative direction can be parametrized by $\thetaJL$ and $\phiJL$.
While $\thetaJL$, the angle between $\Jvec$ and $\Lvec$, depends on the binary intrinsic configuration only, the azimuthal angle $\phiJL$ is measured against a reference axis, $\xhatJ$, which is, in turn, defined against the observer's position (as is $\yhatJ$).
This is made clear in \cref{fig:changes_with_two_perspectives}.
It therefore also transforms in our lensing scenario.
Notice also that, in contrast with $\xhatJ$ and $\yhatJ$, the third direction $\zhatJ$ is attached to the physical total angular momentum, and thus does not change with the observer, \ie. $\zhatJpm=\zhatJ$.
In the same fashion as \cref{eq:cos_thetaJNpm_definition}, by its definition, the change in $\phiJL$ should read
\begin{equation}
    \cos(\phiJLpm) = \Bigl(\bigl(
        \zhatL - (\zhatL\cdot\zhatJ) \, \zhatJ
        \bigr)
    \cdot \xhatJpm
    \Bigr) \,.
    \label{eq:cos_phiJLpm_definition}
\end{equation}
However, while the approach was useful for $\thetaJN \in [0,\pi)$, it would feel incomplete for an azimuthal angle like $\phiJL$ whose range is $[0, 2\pi)$.
Inverting \cref{eq:cos_phiJLpm_definition} by inverting the cosine returns only solutions in $[0,\pi]$.
Alternatively, one may therefore prefer to use the two-argument arctangent and write instead\footnote{
    We write the arguments of $\atantwo$ in the $(y,x)$ ordering.
    Moreover, uncomfortably, the two-argument arctangent typically ranges between $-\pi$ and $\pi$, not from $0$ to $2\pi$, as announced for $\phiorb$.
    We let the reader do the usual conversions.
}
\begin{equation}
    \phiJLpm
    =
    \atantwo\Bigl(
        \bigl(
        \zhatL - (\zhatL\cdot\zhatJ) \, \zhatJ
        \bigr)\cdot \yhatJpm
        ,\,
        \bigl(
        \zhatL - (\zhatL\cdot\zhatJ) \, \zhatJ
        \bigr)\cdot \xhatJpm
    \Bigr) \,,
    \label{eq:phiJLpm_wrt_atan2}
\end{equation}
whose first-order expansion reads
\begin{equation}
    \phiJLpm
    =
    \phiJL
    \pm
    \frac{\openanglepm 2 \sqrt{2} \cos (\philns) \csc (\thetaJN) \sin (\thetalns)}{
        \sqrt{
            \begin{aligned}
                -4 \sin (\philns) & \sin (2 \thetaJN) \sin (2 \thetalns)
                +4 \cos (2 \philns) \sin ^2(\thetaJN) \sin ^2(\thetalns) \\
                - & \cos (2 \thetaJN) (3 \cos (2 \thetalns)+1)-\cos (2 \thetalns)+5
            \end{aligned}
    }}+\bigOof{\openanglepm^2}
    \,.
    \label{eq:phiJLpm_approx}
\end{equation}

\subsection{Phase of coalescence \texorpdfstring{$\phicoal$}{ϕcoal} and reference orbital phase \texorpdfstring{$\phiorb$}{ϕorb}}

We now consider the phase of coalescence, \ie. $\phicoal \defdby \frac{\pi}{2} - \phiorb(t=\tcoal)$.
For the sake of brevity, we do not write out the reference orbital phase $\phiorb$ further (since it is only a shift away).
To express $\phicoal$, we first need the direction of the line of ascending nodes
$\ascnodehat$.
For each virtual observer, it can be expressed thanks to
\begin{equation}
    \ascnodehatpm \propto \Nhatpm \wedge \zhatL \,.
    \label{eq:ascnodehatpm_definition}
\end{equation}

The same remark made about $\phiJL$ applies to $\phicoal$.
The phase of coalescence is an azimuthal angle whose range is $[0, 2\pi)$.
This therefore motivates us to instead consider writing
\begin{equation}
    \phicoalpm = \atantwo\bigl(
        \xhatL \cdot \ascnodehatpm,\,
        - \yhatL \cdot \ascnodehatpm
    \bigr)
    \quad\Leftrightarrow\quad
    \phiorbpm = \atantwo\bigl(
        - \yhatL \cdot \ascnodehatpm,\,
        \xhatL \cdot \ascnodehatpm
    \bigr) \, .
    \label{eq:phicoalpm_wrt_atan2}
\end{equation}
After a series expansion in small opening angle, and assuming that $\Lvec \parallel \Jvec$, one obtains
\begin{equation}
    \phicoalpm  \biggr\rvert_{\Jvec \parallel \LAGNvec}
    =
    \phicoal \pm \frac{2 \openanglepm  \cos (\philns) \csc (\thetaJN)}{\sqrt{2 \cos (2 \philns) \sin ^2(\thetaJN)+\cos (2 \thetaJN)+3}}
    +\bigOof{\openanglepm^2} \,.
    \label{eq:phicoalpm_approx_if_L_and_J_aligned}
\end{equation}

Let us here slip a short comment on the \enquote{hierarchy} between angles.
The lensing plane is defined by the plane containing the source, the lens and the observer.
On the one hand, we observe that if $\incl - \frac{\pi}{2}$ is small with respect to $\philns-\pi$, the lensing plane is basically perpendicular to $\zhat$, and $\inclp \approx \inclm$, \ie. there is virtually no change in the apparent inclination angles.
However, the change seen in $\phicoalpm$ is then maximal.
On the other hand, if it is $\philns-\pi$ that is small compared to $\incl - \frac{\pi}{2}$, then the lensing plane is roughly aligned with $\zhat$, and the opposite remark can be made: the change in the apparent inclination angles $\inclpm$ is maximal, while there is almost no modification of the phase of coalescence $\phicoalpm$.

\subsection{Polarization angle \texorpdfstring{$\psi$}{ψ}}

The definition of the polarization angle $\psi$ is illustrated in \cref{fig:source_J_and_polarization_frames} and detailed in \citet{Isi:2022mbx}.
In short, the polarization angle describes how the observer's frame and coordinate system (those typically being the declination and right ascension) relate to the source frame of the binary.
We observe there that if the observer's direction is moved, the basis vectors of the source frame must be modified as well.

From the same considerations,
\begin{subequations}
    \begin{align}
        \psipm & = \atantwo\bigl(
            \uvec{x}_{W,\pm} \cdot \uvec{y}'_{W,\pm}
            ,\,
            \uvec{x}_{W,\pm} \cdot \uvec{x}'_{W,\pm}
        \bigr)
        \label{eq:psipm_wrt_atan2}
        \\
        & = \atantwo\bigl(
            \uvec{y}_{W,\pm} \cdot (-\uvec{x}'_{W,\pm})
            ,\,
            \uvec{y}_{W,\pm} \cdot \uvec{y}'_{W,\pm}
        \bigr)
        \,.
        \label{eq:psipm_wrt_atan2_alt}
    \end{align}
\end{subequations}
We note also that an angle $\Omega$ is introduced in its definition.
That angle $\Omega$ is usually set to be $\frac{\pi}{2}$, by \ac{lvk}, in which case, the following can be equivalently written
\begin{subequations}
    \begin{align}
        \psipm & = \atantwo\bigl(
            \ascnodehatpm \cdot (-\uvec{x}'_{W,\pm})
            ,\,
            \ascnodehatpm \cdot \uvec{y}'_{W,\pm}
        \bigr)
        \label{eq:psipm_wrt_atan2_with_Omega_set}
        \\
        & = \atantwo\bigl(
            \Lperphatpm \cdot \uvec{y}'_{W,\pm}
            ,\,
            \Lperphatpm \cdot \uvec{x}'_{W,\pm}
        \bigr)
        \,.
        \label{eq:psipm_wrt_atan2_with_Omega_set_alt}
    \end{align}
\end{subequations}
As the explicit expression of the polarization frames is obtained after several layers of rotations (\cref{app:explicit_definition_of_the_basis_vectors}), the formulas for its associated basis $(\uvec{x}_W, \uvec{y}_W, \uvec{z}_W)$ are rather large, and omitted hereafter.
Nonetheless, in the limit of small angles, and if $\Jvec \parallel \LAGNvec$, then the expression for $\psi$ reads
\begin{equation}
    \psipm  \biggr\rvert_{\Jvec \parallel \LAGNvec}
    =
    \psi
    \pm \frac{2 \openanglepm  \cos (\philns) \cot (\thetaJN)}{\sqrt{2 \cos (2 \philns) \sin ^2(\thetaJN)+\cos (2 \thetaJN)+3}}
    + \bigOof{\openanglepm^2}
    \,.
    \label{eq:psipm_approx_if_L_and_J_aligned}
\end{equation}

This ends this section, where we have shown explicitly the transformation rules for the inclination angle $\thetaJN$, the angle $\phiJL$, the phase of coalescence $\phicoal$, and the polarization angle $\psi$: \cref{eq:thetaJNpm_approx_if_L_and_J_aligned,eq:phiJLpm_approx,eq:phicoalpm_approx_if_L_and_J_aligned,eq:psipm_approx_if_L_and_J_aligned}, respectively.
All these transformations are applicable to the lensing considered here: lensing of a \ac{cbc}, behind a nearby supermassive black hole.

\section{Doppler shift induced by the orbital velocity}
\label{sec:Doppler_shift}

At the approximation level that we are investigating, the last step that is required to fully describe the lensed waveforms is to quantify the Doppler shift due to the orbital rotation velocity $\vorbit$ (\cref{eq:vorbit_definition}).

By construction (\cref{fig:source_J_and_polarization_frames}), the supermassive black hole is located at
\begin{equation}
    \vec{r}_\text{SMBH} = \rorbit \cdot \uvec{r}_\text{SMBH} \,,
    \label{eq:rSMBH_definition}
\end{equation}
where
\begin{equation}
    \uvec{r}_\text{SMBH} =
        \cos (\philns) \sin (\thetalns) \cdot \xhatJ
        + \sin (\philns) \sin (\thetalns) \cdot \yhatJ
        + \cos (\thetalns) \cdot \zhatJ \,.
    \label{eq:rSMBH_hat_definition}
\end{equation}
By definition of $\phimis$, the spin direction of the \ac{agn} can be written as
\begin{equation}
    \LAGNhat =
    R_{\uvec{r}_\text{SMBH}} (\phimis)
    \cdot R_{\zhatJ} (\philns)
    \cdot R_{\yhatJ} (\thetalns-\sfrac{\pi}{2})
    \cdot \zhatJ
    \,,
    \label{eq:LAGN_definition}
\end{equation}
where $R_{\vec{u}}(\theta)$ is the rotation matrix around $\vec{u}$ by an angle $\theta$.
Accordingly, by the remark on the orthogonality given in \cref{sec:orbital_total_angular_momentum_and_polarization_frames}, the orbital velocity $\vorbitvec$ reads
\begin{align}
    \vorbitvec & = \vorbit \cdot \left( \LAGNhat \wedge \uvec{r}_\text{SMBH} \right) \,,
    \label{eq:vorbitvec_definition}
\end{align}
where \enquote{$\wedge$} denotes the usual cross product.
From there, and with the previously defined observer directions (\cref{eq:virtual_observer_directions_definition}), the apparent velocity along each line of sight is
\begin{equation}
    \vorbitpm = \vorbitvec \cdot \Nhatpm \,.
    \label{eq:vorbitpm_compact_definition}
\end{equation}
The resulting expression is large and is thus not reproduced here.
However, a few remarks can be made.
First, if $\philns = \frac{\pi}{2}$ and $\phimis=0$, there is virtually no Doppler shift.
Indeed, in that case, the lensing plane is perpendicular to the orbital plane of the binary revolving around the central supermassive black hole, and $\vorbitvec \perp \Nhatpm$.
Conversely, if $\thetalns = \frac{\pi}{2}$ and $\phimis=0$, these two planes are coplanar, and the impact of the Doppler shift is maximal.
If $\phimis=0$, at first order in small angles, we can write
\begin{equation}
    \frac{\vorbitpm}{\vorbit}
    =
    \cos (\philns) \sin (\thetaJN)
    \left(
        1
        \pm
        \openanglepm
        \frac{  \sin (\philns) \sin (\thetaJN) \sin (\thetalns)+\cos (\thetaJN) \cos (\thetalns)}{\sqrt{\denomInJFrame}}+\bigOof{\openanglepm^2}
    \right)
    \,,
    \label{eq:vorbitpm_approx_with_phimis_null}
\end{equation}
The difference $\Delta \vorbitpm \defdby \vorbitp - \vorbitm$ between the two signals, under the same assumptions, is thus
\begin{equation}
    \frac{\Delta \vorbitpm}{\vorbit}
    =
    \openangle \cos (\philns) \sin (\thetaJN) \frac{\sin (\philns) \sin (\thetaJN) \sin (\thetalns)+\cos (\thetaJN) \cos (\thetalns)}{\sqrt{\denomInJFrame}}+\bigOof{\openangle^2} \,.
    \label{eq:difference_in_vorbitpm_approx_with_phimis_null}
\end{equation}
Notice how the result only depends on the ratio of $\MSMBH$ and $\rorbit$ (via $\vorbit$, \cref{eq:vorbit_definition}).
Therefore, any information on the redshift difference will be tied to it.

To get a sense of the amplitude of the Doppler shift, assume $\phimis=0$ and set the gravitational-wave source (almost) exactly behind the supermassive black hole, \ie. $\thetaJN=\frac{\pi}{2}$, $\thetalns=\frac{\pi}{2}$ and $\philns \to \frac{\pi}{2}$.
Then, \cref{eq:difference_in_vorbitpm_approx_with_phimis_null} reduces to (be careful to consider the dependencies hidden in $\denomInJFrame$ in doing so)
\begin{subequations}
    \begin{align}
        \bigl| \Delta v_\pm \bigr| & = \vorbit \cdot \openangle + \bigOof{\openangle^2}
        \label{eq:difference_in_vorbitpm_approx_on_LOS} \\
        & \approx 2 \, \frac{\RS}{\rorbit} = 4 \, \frac{G}{c^2} \frac{\MSMBH}{\rorbit}
        \label{eq:difference_in_vorbitpm_approx_on_LOS_wrt_RS}
    \,.
    \end{align}
\end{subequations}
where \cref{eq:vorbit_definition,eq:openanglepm_approx_on_LOS} were used in the second line.
This last result can be used for estimating the order of magnitude of the Doppler shift with respect to the mass of the lensing supermassive black hole, and the distance of the source binary from it.
Again, we observe that it is the ratio of the two that matters.

The global impact of all the changes due to the viewing perspectives and the corresponding Doppler shift is represented in \cref{fig:lensed_waveforms}.
All these changes are fundamentally due to the emission geometry of the system.

\section{Discussion}
\label{sec:discussion}

Now that we have made explicit the formulas for describing the gravitational-wave parameters obtained in the case of lensing of a gravitational-wave source by a supermassive black hole in the geometrical limit (\cref{eq:thetaJNpm_approx,eq:phiJLpm_approx,eq:phicoalpm_approx_if_L_and_J_aligned,eq:psipm_approx_if_L_and_J_aligned,eq:difference_in_vorbitpm_approx_with_phimis_null}), we address some possible concern the reader may hold, before discussing the results.

So far, we have not incorporated any \emph{explicit} time dependence in any of the computations.
However, since the AGN-CBC system is a dynamical system, this would be necessary to obtain a more faithful description of the signal's parameters.
To describe how the parameters change during the course of the orbital motion, all parameters of the AGN-CBC system can be promoted to dynamical functions.
This change thus echoes onto the transformed binary parameters in the same go, as well as the lensing parameters (\eg., magnification, time delay, phase shift).
The impact of the orbital motion of the binary on other parameters grows with its proximity to the lensing supermassive black hole (since $\openangle$ increases; see \cref{app:asymptotic_expressions_for_an_infinitely_far_away_observer,fig:quantities_at_infinity}).
Nonetheless, for \acp{bbh}, because the signal duration $\signalduration$ is short relative to the revolution period, that impact is expected to remain small in any reasonable parameter space (as the calculation on the probability of truncation in \cref{app:probability_of_truncation} shows en passant).

We now address two other side questions.
Firstly, when picturing a binary orbiting a supermassive black hole, the former can enter and exit that region, while travelling on its circular path around the said supermassive black hole.
It is however unlikely for the signal to be truncated by such an occurrence (\cref{app:probability_of_truncation}), for $\MSMBH \geq \qty{1e6}{\solarmass}$ (see \cref{fig:quantities_at_infinity}).
Secondly, what are the chances for the two signals (resulting from lensing) to overlap with each other?
The gravitational-wave signal from a \ac{bbh} typically lasts \qty{1}{\second} or less, while in most reasonable parameter ranges the arrival time difference can reach up to a few minutes, leaning towards longer time differences (see \cref{fig:quantities_at_infinity}).
Again, it is therefore unlikely for the signal to overlap with itself (\cref{app:probability_of_signal_overlap}).

These considerations may not hold for \acp{bns}, for which the longer inspiral ($\bigOof{\unit{\minute}}$) could require incorporating the time variation to describe these parameters.
In that case, the chances of truncation or signal overlap increase proportionally.
The change remains small with heavier \acp{smbh} ($\gtrapprox\qty{1e7}{\solarmass}$).
Nevertheless, the formalism here would still be applicable, as the two signals could be modelled then using the standard amplification factor approach~\cite{Takahashi:2003ix,Lai:2018rto}, with the two images having the modified viewing angles and Doppler shifts derived here.

To achieve a full description of the system, several other considerations can be raised, and we answer some of these concerns.
First, the central supermassive black hole is not a static point mass, but is better described by a near-extremal rotating Kerr black hole, which means that the nearby spacetime would be frame-dragged. %
To test our working hypothesis, we have qualitatively explored the system using publicly available numerical geodesic finders, namely Gyoto~\cite{Vincent:2011wz,Aimar:2023vcs,gyoto_doi} and \texttt{kgeo}~\cite{Gralla:2019ceu,chael2022_kgeo}.
These qualitative explorations have shown that the thin-lens approximation was a good model to approach the problem.
Furthermore, we have neglected the mass of the accretion disk, which, in principle, can be large. %
However, its impact is arguably limited and, by making the mass of the lens heavier, should make the effect described in this work more pronounced.
At first order, the accretion disk's mass could thus bias the distance between the source and the lensing supermassive black hole.

Combining the effects described in \cref{sec:geometrical_transformation_rules_of_angles,sec:Doppler_shift} enables one to write down the parameters of the waveforms of a gravitational-wave source orbiting and lensed by the \ac{smbh} of an \ac{agn}.
\Cref{fig:lensed_waveforms} illustrates what this could look like in practice.
It reveals that, in these systems, the impact of the two different perspectives on the same event can effectively shift the phase of the waveforms, amplify/diminish their amplitude and change their frequency.
If \acp{agn} are a significant formation channel for gravitational-wave binary sources, these remarks could have consequences on the search and detection of such lensed gravitational-wave signals.
Furthermore, this also has the potential of bringing in new and unique insights into binary mergers and \ac{agn} systems.
In addition to their magnifications, time delay and Morse phase, the two lensed signals would thus bring \num{8} angles ($\thetaJN$, $\phiJL$, $\phiorb$ and $\psi$, for each image) and \num{2} Doppler shifts, all characterized by only \num{5} parameters (\ie. $\rorbit$, $\thetalns$, $\philns$, $\phimis$ and $\MSMBH$).
Subsequent works should determine how well these underlying parameters can be recovered in realistic scenarios.

\begin{figure}[H]
    \centering
    \input{tikz/lensed_waveforms.tex}
    \caption{
        Gravitational-wave waveform lensed by the \ac{smbh} of an \ac{agn}, obtained after the transformations described in \cref{sec:geometrical_transformation_rules_of_angles,sec:Doppler_shift} have been applied.
        The top panel schematically pictures the timeline of gravitational lensing in the point-mass lens approximation.
        As for more traditional lensing scenarios, gravitational lensing translates a gravitational-wave signal ($h_{\times, \neg\text{lns}}$) into two duplicated versions that are delayed and magnified (or demagnified and phase shifted) ($h_{\times,\pm,\text{no effect}}$), thus yielding two distinct so-called \enquote{images} denoted here by $+$ and $-$.
        The impact of the effects (magnifications and the Morse phase shift) is shown in the panel below.
        The following intermediary panel individually adds the impact of either the geometrical transformations of \cref{sec:geometrical_transformation_rules_of_angles} (on the inclination $\thetaJN$, angle $\phiJL$, orbital phase $\phiorb$ and polarization angle $\psi$) or the Doppler shift of \cref{sec:Doppler_shift}\,---but not both.
        It is in the bottom panel that both effects are combined and the gravitational-wave waveform expected from gravitational lensing in an \ac{agn} is finally exhibited.
        Notice how these effects yield a clearly different waveform compared to what the original waveform (second to top panel) looks like.
        In this example, the combination of the effects even shifts the phase of the $-$ image enough to make it temporarily look like the unlensed waveform.
        The PyCBC Python package~\cite{pycbc,Ramos-Buades:2023ehm} was used to generate these waveforms, and all original parameters are compiled in \cref{app:Example_waveform_parameters}.
    }
    \label{fig:lensed_waveforms}
\end{figure}

\acknowledgments

PM, SHWL, OAH acknowledge support by grants from the Research Grants Council of Hong Kong (Project No. CUHK 14304622, 14307923, and 14307724), the start-up grant from the Chinese University of Hong Kong, and the Direct Grant for Research from the Research Committee of The Chinese University of Hong Kong.
The authors acknowledge the help from Aditya Sharma and Helena Ubach in discussing the problem and sharing information, and Helena Ubach additionally for reviewing the paper during the LIGO pnp process.

\appendix

\section{Definition of the basis vectors}
\label{app:explicit_definition_of_the_basis_vectors}

In what follows we give an expression of the basis vectors, as they are defined via \cref{fig:source_J_and_polarization_frames,fig:lensing_plane_frame}.
We adopt here the basis vectors associated with the $\Jvec$-frame, \ie. $(\xhatJ, \yhatJ, \zhatJ)$ for the unlensed observer.
We write the usual cross product with \enquote{$\wedge$}, and $R_{\vec{u}}(\omega)$ denotes the rotation matrix around the direction $\vec{u}$ by an angle $\omega$.

\subsection{Positions of the source, the observer, and the lens}

We decide to place the source at the origin ($\vec{r}_\text{src} = (0, 0, 0)$), and the positions of the observer and the lens are straightforward to express.
The (unlensed) observer is in the direction
\begin{equation}
    \Nhat =
    \begin{pmatrix}
        0 \\
        \sin(\thetaJN) \\
        \cos(\thetaJN)
    \end{pmatrix}\,,
\end{equation}
and it is thus located at
\begin{equation}
    \robsvec = \Ds \cdot \Nhat \,.
\end{equation}
In the case of lensing, the apparent observer direction is given in \cref{eq:virtual_observer_directions_definition}.
The lens is situated using $\rorbit$, $\thetalns$ and $\philns$, and its vector position reads
\begin{equation}
    \rlnsvec = \rorbit \cdot
    \begin{pmatrix}
        \sin(\thetalns) \cos(\philns) \\
        \sin(\thetalns) \sin(\philns) \\
        \cos(\thetalns)
    \end{pmatrix} \,.
\end{equation}
Once the three bodies are positioned (and given a mass $\MSMBH$ and rotation direction via $\LAGNvec$), the lensing system is fully determined.

\subsection{Basis vectors of the orbital or source frame}

Let us now define the basis vectors of the $\Lvec$-based orbital frame or source frame, \ie. $(\xhatL, \yhatL, \zhatL)$.
With respect to the frame associated with $\Jvec$, and following \cref{fig:source_J_and_polarization_frames}, we express $\zhatL$ using the angles $\phiJL$ and $\thetaJL$ as
\begin{equation}
    \zhatL =
    \begin{pmatrix}
        \cos(\phiJL) \sin(\thetaJL) \\
        \sin(\phiJL) \sin(\thetaJL) \\
        \cos(\thetaJL)
    \end{pmatrix} \,,
\end{equation}
where we have introduced the angle $\thetaJL$, \ie. the angle between the total and the orbital angular momenta.
This angle is not a free parameter, and \cref{app:angle_theta_JL} describes how to compute it via the gravitational-wave parameters.
Now, the line of ascending nodes is simply
\begin{equation}
    \ascnodehat \propto \Nhat \wedge \zhatL \,,
\end{equation}
(and $\Lperphat \defdby \ascnodehat \wedge \Nhat$).
It enables us to write
\begin{equation}
    \xhatL = R_{\zhatL}(\phiorb) \cdot \ascnodehat \,,
    \label{eq:explicit_expression_for_xhatL}
\end{equation}
and we complete the triad via $\yhatL = \zhatL \wedge \xhatL$.
This frame is anchored in physical quantities and is independent of lensing; its expression with respect to another frame is not necessarily so.

\subsection{Basis vectors of the polarization frame}

We finally define the so-called polarization frame.
By construction (\cref{fig:source_J_and_polarization_frames}), one of its directions is fixed by the observer's direction, \ie.
\begin{equation}
    \uvec{z}_W = \uvec{z}_W' = \Nhat \,.
\end{equation}
Then, we get $\uvec{x}_W$ with respect to the line of ascending nodes $\ascnodehat$ after a rotation as
\begin{equation}
    \uvec{x}_W = R_{\Nhat}(-\Omega) \cdot \ascnodehat \, ,
\end{equation}
and the third basis vector just completes the triad: $\uvec{y}_W = \uvec{z}_W \wedge \uvec{x}_W$.

As for its primed counterpart, we had $\uvec{z}_W = \uvec{z}_W' = \Nhat$ and another component is just obtained in the same way, after an extra rotation.
Explicitly, we write
\begin{subequations}
    \begin{align}
        \uvec{x}_W'
            & = R_{\Nhat}(-\psi) \cdot \uvec{x}_W \\
            & = R_{\Nhat}(-\Omega-\psi) \cdot \ascnodehat \, ,
    \end{align}
\end{subequations}
and $\uvec{y}_W'$ is obtained as before (or by rotating $\Lperphat$ instead of $\ascnodehat$).

\subsection{Basis vectors of the lensing frame}
\label{app:basis_vectors_of_the_lensing_frame}

The lensing frame helps us describe the lensing and is thus also independent of the apparent observer's positions (since it defines them).
We choose
\begin{equation}
    \xhatlns \defdby \Nhat \,
\end{equation}
and we choose $\zhatlns$ to be perpendicular to the lensing plane.
Explicitly, we write it as
\begin{equation}
    \zhatlns \propto \left(
        \rlnsvec - \vec{r}_\text{src}
    \right) \wedge \Nhat \,
\end{equation}
so that the last vector basis $\yhatlns$ points away from the line of sight (that is up in \cref{fig:lensing_plane_frame}).

\subsection{Frames that change with the observer's direction}

Crucially, the polarization frame and the frame associated with the total angular momentum $\Jvec$ depend on the apparent observer's direction (\cref{eq:virtual_observer_directions_definition}).
Indeed, depending on its direction, we have that
\begin{equation}
    \xhatJpm \propto \Nhatpm \wedge \zhatJ \,,
\end{equation}
and similarly for $\yhatJpm$.
The line of ascending nodes also needs to be adapted (\cref{eq:ascnodehatpm_definition}), which, in turn, impacts the polarization frame.
For example, we get
\begin{equation}
    \uvec{x}_{W,\pm} = R_{\Nhat}(-\Omega) \cdot \ascnodehatpm\,.
\end{equation}
Noticeably, this is not how the other primed frame is transformed.
Instead, since the primed frame is associated with the observer\,---and the polarization angle $\psi$ relates the two---, we must write
\begin{equation}
    \uvec{x}_{W,\pm}' = R_{\zhatlns}(\pm\openanglepm) \cdot \uvec{x}_{W,\pm} \,.
\end{equation}
That explains why the polarization angle $\psi$ also changes.
Now, one has explicit steps to reproduce the results compiled in this work.

\section{Orientation angles that change with the observer's direction}
\label{app:highglighting_the_angles_that_change}

\Cref{fig:angle_definitions_in_isolation} highlights the $4$ angles that change with the observer's position.
These are the four gravitational-wave angles that are changing because of the geometry specific to lensing in an \ac{agn}.
The same angles are also defined in \cref{fig:source_J_and_polarization_frames}, but in a single diagram.

\begin{figure}[H]
    \centering
    \input{tikz/angle_definitions_in_isolation.tex}
    \caption{
        Diagrams exhibiting exclusively how $\thetaJN$, $\phiJL$, $\phiorb$ (or $\phicoal$), and $\psi$ are defined (see also \cref{fig:source_J_and_polarization_frames}).
        The leftmost (a) panel shows the inclination angle $\thetaJN$ of the total angular momentum $\Jvec$ with respect to the observer's direction $\Nhat$.
        The middle left (b) panel pictures the plane perpendicular to $\Jvec$ to exhibit how the azimuthal angle $\phiJL$ is defined between $\xhatJ$ and the (projected) orbital angular momentum $\Lvec$.
        The middle right (c) panel focuses on the orbital phase $\phiorb$, and its complementary angle $\phicoal$, defined in the plane perpendicular to the orbital angular momentum $\Lvec$.
        It describes the angle between $\xhatL$ and the observer's direction $\Nhat$, or the line of ascending nodes $\ascnodehat$.
        The rightmost (d) panel provide a \enquote{sky projection} (\ie. the plane perpendicular observer's direction $\Nhat$), and thus exhibits the angle $\Omega$, as well as the polarization angle $\psi$.
        By setting $\Omega = \frac{\pi}{2}$, the \ac{lvk} Collaboration make it so that\,---when projected so---\,$\uvec{y}_W$ coincides with the line of ascending nodes $\ascnodehat$, and $\uvec{x}_W$ coincides with $\zhatL$.
    }
    \label{fig:angle_definitions_in_isolation}
\end{figure}

\section{Parameters introduced in an AGN-CBC system}
\label{app:highglighting_the_new_parameters}

\Cref{fig:new_parameters_only} highlights the $5$ parameters necessary to fully determine the changes described in the current work, \ie. the position of the lensing \ac{smbh} via $\rorbit$, $\thetalns$, and $\philns$, the mass $\MSMBH$ of the lensing \ac{smbh}, and the misalignment angle $\phimis$.
These were introduced in \cref{sec:system_setup,fig:source_J_and_polarization_frames}.

\begin{figure}[H]
    \centering
    \input{tikz/new_parameters_only.tex}
    \label{fig:new_parameters_only}
    \caption{
        Parameters introduced to determine the lensing system as modelled in this work (in black).
        The lensing \ac{smbh} is located in spherical coordinates with respect to the frame associated with the total angular momentum $\Jvec$, \ie. via the radial distance $\rorbit$, the zenith angle $\thetalns$, and the azimuthal angle $\philns$.
        The mass $\MSMBH$ of the \ac{smbh} is also necessary, or, equivalently, its Schwarzschild radius $\RS$ (\cref{eq:Schwarzschild_radius_definition}).
        Finally, the misalignment angle $\phimis$ intervenes in the Doppler shift (\cref{eq:LAGN_definition}).
    }
\end{figure}

\section{Example waveform generation parameters}
\label{app:Example_waveform_parameters}

\begin{table}[H]
    \centering
    \caption{
        All parameters used to generate the waveforms exhibited in \cref{fig:lensed_waveforms}.
        Notice that the masses $m_1$ and $m_2$ are given in the source frame, and the luminosity distance was inferred from the chosen redshift $z$.
    }
    \begin{tabular}{lcc}
        \toprule
        \multicolumn{3}{c}{Intrinsic gravitational-wave parameters} \\
        \cmidrule(lr){1-3}
        \multirow{2}*{Masses in the source frame}
            & $m_1$ & \qty{36}{\solarmass} \\
            & $m_2$ & \qty{29}{\solarmass} \\
        \multirow{2}*{Dimensionless spin magnitudes} & $a_1$ & \num{0.4} \\
        & $a_2$ & \num{0.3} \\
        \multirow{2}*{Polar angle between spin and orbital angular momentum $\Lvec$}
            & Tilt $1$ & \qty{0.5}{\radian} \\
            & Tilt $2$ & \qty{1.0}{\radian} \\
        Azimuthal angle between the two spin vectors
            & $\phi_{12}$ & \qty{1.7}{\radian} \\
        Azimuthal angle between the total angular momentum $\Jvec$
            & \multirow{2}*{$\phiJL$} & \multirow{2}*{\qty{0.3}{\radian}} \\
        \quad and the orbital one $\Lvec$ & & \\
        \cmidrule{1-3}
        \multicolumn{3}{c}{Extrinsic gravitational-wave parameters} \\
        \cmidrule(lr){1-3}
        Redshift of the \ac{agn} system
            & $z$ & \num{0.1} \\
        Inclination angle between the total angular momentum $\Jvec$
        & \multirow{2}*{$\thetaJN$} & \multirow{2}*{\qty[parse-numbers=false]{0.99 \cdot \frac{\pi}{2}}{\radian}} \\
        \quad and the line of sight ($\Nhat$) & & \\
        Orbital phase at the reference frequency
            & $\Phi$ & \qty{1.3}{\radian} \\
        Right ascension
            & $\ra$ & \qty{1.375}{\radian} \\
        Declination
            & $\dec$ & \qty{1.2108}{\radian} \\
        Time of the merger
            & $t_\text{geo}$ & \qty{1126259642.413}{\second} \\
        Gravitational-wave polarization
            & $\psi$ & \qty{2.659}{\radian} \\
        \cmidrule{1-3}
        \multicolumn{3}{c}{\Ac{agn} lensing parameters} \\
        \cmidrule(lr){1-3}
        Dimensionless source position
            & $y$ & \num{0.5} \\
        Orbital radius of the source
            & $\rorbit$ & \qty{100}{\RSunit} \\
        Angle placing the source
            & $\psi_\text{lns}$ & \qty[parse-numbers=false]{\frac{\pi}{6}}{\radian} \\
        Mass of the lensing \ac{smbh}
            & $\MSMBH$ & \qty{1e6}{\solarmass} \\
        Misalignment angle (\cref{fig:source_J_and_polarization_frames})
            & $\phimis$ & \qty{0.1}{\radian} \\
        \cmidrule{1-3}
        \multicolumn{3}{c}{Numerical parameters} \\
        \cmidrule(lr){1-3}
            \multicolumn{2}{l}{Approximant} & \texttt{SEOBNRv5PHM}~\cite{Ramos-Buades:2023ehm} \\
            Reference frequency & $f_\text{ref}$ & \qty{20}{\hertz} \\
        \bottomrule
    \end{tabular}
\end{table}

\section{Asymptotic expressions for an infinitely far-away observer}
\label{app:asymptotic_expressions_for_an_infinitely_far_away_observer}

One can show that in the limit where the observer is placed far away from the source and the lens\,---\ie. $\Dls \ll \Ds \approx \Dl$---, the opening angles $\openanglepm$ can be expressed independently of $\Dl$ and $\Ds$ (\cref{app:gamma_pm_for_a_far_away_observer}).
Similar formulas can be obtained to describe the arrival time difference and the magnification factors (\cref{app:arrival_time_difference_for_a_far_away_observer,eq:magnification_factors_for_a_far_away_observer}).

\begin{figure}[h]
    \centering
    \tikzsetnextfilename{quantities_at_infinity}
\begin{tikzpicture}
    \begin{groupplot}[
        group style={
            group name=quantities at infinity,
            group size=3 by 1,
            horizontal sep=2.2cm,
        },
        width=0.205\textwidth,
        height=0.205\textwidth,
        ylabel style={yshift=-1mm,},
        enlargelimits=false,
        axis on top,
        colorbar style={
        },
        title style={
            align=center,
            text width=0.22\textwidth,
        },
    ]
        \nextgroupplot[
            title={Total opening angle \\ $\openangle \equiv \openanglep + \openanglem$},
            point meta min=1.6204815970290225,
            point meta max=98.38459900738053,
            ylabel={$r_\text{lns} / R_\text{Sch}$ [\phantom{m}]},
            ymode=log,
            ymin=1e1,
            ymax=1e4,
            xmode=log,
            xmin=1e-2,
            xmax=1e1,
            colormap/viridis,
            colorbar,
            colorbar style={
                width=2mm,
                xshift=-5mm,
                xticklabel pos=right,
                xlabel={[\unit{\degree}]},
            },
        ]
            \addplot graphics[
                xmin=1e-2, xmax=1e1,
                ymin=1e1, ymax=1e4,
            ] {pictures/total_opening_angle.png};
            \node[above, rotate={0}] at (axis cs:0.1, 4000) {\tiny\color{White}\ang{2.5}};
            \node[above, rotate={0}] at (axis cs:0.1, 1000) {\tiny\color{White}\ang{5}};
            \node[above, rotate={0}] at (axis cs:0.1, 250) {\tiny\color{White}\ang{10}};
            \node[above, rotate={0}] at (axis cs:0.1, 65) {\tiny\color{White}\ang{20}};
            \node[above, rotate={0}] at (axis cs:0.1, 13) {\tiny\color{White}\ang{40}};
            \node[above, rotate={55}] at (axis cs:5, 30) {\tiny\color{White}\ang{60}};
            \node[above, rotate={55}] at (axis cs:7, 20) {\tiny\color{White}\ang{80}};
            \addplot[white, no marks, thin, unbounded coords=jump, forget plot,]
                table[col sep=comma] {csv_tables/total_opening_angle_contours.csv};
            \addplot[mark=none, densely dotted, White, forget plot,] coordinates {(1,1e1) (1,1e4)};
            \node[above, rotate={90}] at (axis cs:1,1150) {\color{White}$y = 1$};

        \nextgroupplot[
            title={Arrival time difference \\ $\Delta t_\mathrm{arrival}$},
            point meta min=-0.4044751496834907,
            point meta max=9.039685547489919,
            ylabel={$M_\text{lns}$ [$M_\odot$]},
            ymode=log,
            ymin=1e6,
            ymax=1e9,
            xmode=log,
            xmin=1e-2,
            xmax=1e1,
            colormap/viridis,
            colorbar,
            colorbar style={
                width=2mm,
                xshift=-5mm,
                xticklabel pos=right,
                xlabel={[\unit{\second}]},
                yticklabel={$10^{\pgfmathprintnumber{\tick}}$},
            },
        ]
            \addplot graphics[
                xmin=1e-2, xmax=1e1,
                ymin=1e6, ymax=1e9,
            ] {pictures/time_delay_difference.png};
            \addplot[mark=none, densely dotted, White, forget plot,] coordinates {(1,1e6) (1,1e9)};
            \node[above, rotate={90}] at (axis cs:1,14000000) {\color{White}$y = 1$};
            \addplot[white, no marks, thin, unbounded coords=jump]
                table[col sep=comma] {csv_tables/time_delay_difference_contours.csv};
            \node[above, rotate={-48}] at (axis cs:0.015,1500000) {\tiny\color{White}\qty{1}{\second}};
            \node[above, rotate={-48}] at (axis cs:0.04,5300000) {\tiny\color{White}\qty{1e1}{\second}};
            \node[above, rotate={-48}] at (axis cs:0.1,22500000) {\tiny\color{White}\qty{1e2}{\second}};
            \node[above, rotate={-48}] at (axis cs:0.2,110000000) {\tiny\color{White}\qty{1e3}{\second}};
            \node[above, rotate={-48}] at (axis cs:1.7,130000000) {\tiny\color{White}\qty{1e4}{\second}};
            \node[below, rotate={-48}] at (axis cs:6,550000000) {\tiny\color{White}\qty{1e5}{\second}};

        \nextgroupplot[
            title={Magnifications \\ $\mu_+$ \& $\mu_-$},
            xmode=log,
            xmin=1e-2,
            xmax=1e1,
            ymode=log,
            ymin=1e-3,
            ymax=1e2,
            legend style={
                at={(0.1,0.1)},
                anchor=south west,
            },
        ]
            \addplot[mark=none, densely dotted, Grey, forget plot,] coordinates {(1e-2,1) (1e1,1)};
            \addplot[mark=none, densely dotted, Grey, forget plot,] coordinates {(1,1e-3) (1,1e2)};
            \addplot[
                no marks,
                Tomato,
                thick,
            ]
                table[
                col sep=comma,
                x=y,
                y=magnification_plus,
            ] {csv_tables/magnification_plus_and_minus.csv};
            \addlegendentry{$\mu_+$}
            \addplot[
                no marks,
                CornflowerBlue,
                densely dashed,
                thick,
            ]
                table[
                col sep=comma,
                x=y,
                y=magnification_minus,
            ] {csv_tables/magnification_plus_and_minus.csv};
            \addlegendentry{$\mu_-$}
            \node[below] at (axis cs:0.04,1) {\color{Grey}$\mu_\pm = 1$};
            \node[below, rotate={90}] at (axis cs:1,10) {\color{Grey}$y = 1$};
    \end{groupplot}
    \coordinate (toplabelx) at ($(quantities at infinity c1r1.west)!0.5!(quantities at infinity c3r1.east)$);
    \node[below, yshift=-5mm] at (toplabelx |- quantities at infinity c1r1.south) {Dimensionless image position $y$ [\phantom{m}]};
\end{tikzpicture}
    \caption{
        From left to right, opening angle $\openangle$, arrival time difference $\Delta t_\text{arrival}$ and magnification factors $\mu_\pm$ computed according to \cref{eq:gamma_pm_explicit_for_appendix,eq:asymptotic_arrival_time_difference,eq:magnification_factors_for_a_far_away_observer}, that is in the limit where the observer is placed far away from the whole system (source and lens) ($\Dls \ll \Ds \approx \Dl$).
        The leftmost panel exhibits the total opening angle $\openangle \defdby \openanglep+\openanglem$ in degrees against the orbital radius $\rorbit$ (in Schwarzschild radii $\RS$) and the dimensionless source position $y \defdby \eta \cdot \RE^{-1}$.
        The central panel shows the arrival time difference $\Delta t_\text{arrival}$ in seconds against the mass of the lens $\MSMBH$ and, again, the dimensionless source position $y$.
        The final rightmost panel shows the magnifications $\mu_\pm$ of each image against the dimensionless parameter $y$, the only parameter they depend on.
        }
        \label{fig:quantities_at_infinity}
    \end{figure}
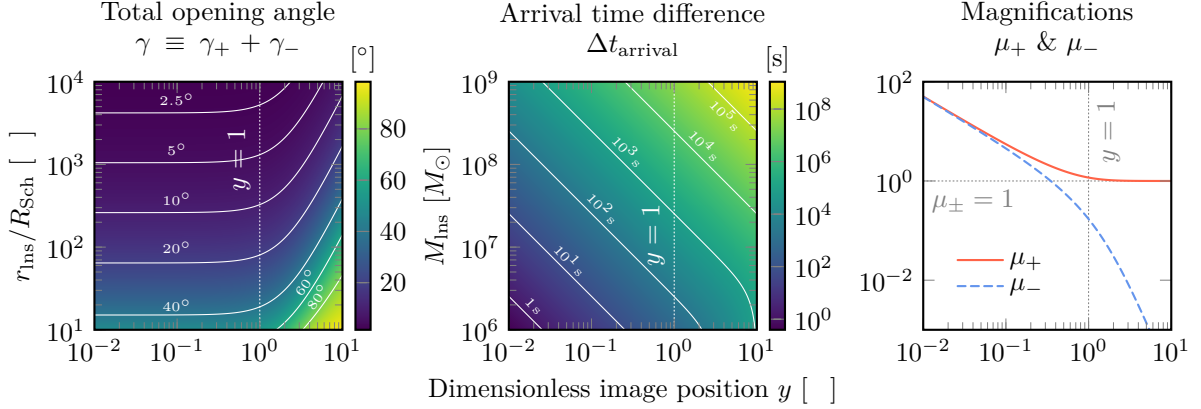

\subsection{The opening angle \texorpdfstring{$\openanglepm$}{γ±}}
\label{app:gamma_pm_for_a_far_away_observer}
Without loss of generality, say the observer is at $(0,0,-\Dl)$, while the lens is at $(0,0,0)$, and the source at $(0,y,z)$ (with $z>0$).

The opening angles $\openanglepm$ (\cref{eq:openanglepm_definition}) then read
\begin{equation}
    \openanglepm = \arctan\left(
        \frac{\Dl \tan(\theta_\pm) \mp \Ds \tan(\beta)}{\Dls}
    \right)
        \pm \beta \,,
    \label{eq:gamma_pm_explicit_for_appendix}
\end{equation}
and one can express $\beta$ as
\begin{equation}
    \beta = \arctan\left(\frac{y}{\Dl+z}\right)\,.
\end{equation}
In the limit where $\Dl \to \infty$, \cref{eq:gamma_pm_explicit_for_appendix} then reduces to
\begin{subequations}
    \begin{align}
        \openanglepm & \approx \arctan \left(
            \frac{1}{2 \Dls} \left(
                \sqrt{\eta^2 + 8 \RS \Dls}
                \mp \eta
            \right)
        \right) \\
        & \approx \arctan\left(
            \sqrt{\frac{\RS}{2 \Dls}} \left(
                \sqrt{y^2 + 4} \mp y
            \right)
        \right) \,,
    \end{align}
\end{subequations}
where $\eta$ is the source position in the lensing plane (see \cref{fig:lensing_plane_frame}).
This expression can be used as a good approximation in the case of an \ac{agn} system, where $\Dls \ll \Ds \approx \Dl$.

\subsection{Arrival time difference}
\label{app:arrival_time_difference_for_a_far_away_observer}

Let us first define the pseudo-angle $\theta_{\pm,\infty}$ as
\begin{equation}
    \theta_{\pm,\infty} \defdby \frac{
        1
    }{
        \sqrt{4 \Dls}
    }
    \left(
        \sqrt{8 \Dls + \frac{b^2}{\RS}} \pm \frac{|b|}{\sqrt{\RS}}
    \right) \,.
\end{equation}
For a point mass, the difference $\Delta t_\pm$ between the emission and arrival times due to lensing can be approximately expressed as \cite{poissonGravityNewtonianPostNewtonian2014}
\begin{align}
    \Delta t_\pm & =
    \frac{1}{c} \left| \robsvec - \vec{r}_\text{src} \right|
    +
    \frac{\RS}{c} \ln\left(
        \frac{
            \left(
                \robs + \robsvec \cdot \uvec{n}_\pm
            \right)
            \left(
                r_\text{src} - \vec{r}_\text{src} \cdot \uvec{n}_\pm
            \right)
        }{
            \left(
                \vec{r}_\text{src} - (\vec{r}_\text{src} \cdot \uvec{n}_\pm) \uvec{n}_\pm
            \right)^2
        }
    \right)
    + \bigOof{c^{-5}} \nonumber \\
    & \approx
    \frac{1}{c} \left| \robsvec - \vec{r}_\text{src} \right|
    +
    \frac{\RS}{c} \ln\left(
        \frac{
            2 \, \robs
            \left(
                r_\text{src} - \vec{r}_\text{src} \cdot \uvec{n}_\pm
            \right)
        }{
            \left(
                \vec{r}_\text{src} - (\vec{r}_\text{src} \cdot \uvec{n}_\pm) \uvec{n}_\pm
            \right)^2
        }
    \right)
    \nonumber \\
    & =
    \frac{1}{c} \left| \robsvec - \vec{r}_\text{src} \right|
    +
    \frac{\RS}{c} \ln\left(
        2 \, \robs
        \frac{
            r_\text{src} - \vec{r}_\text{src} \cdot \uvec{n}_\pm
        }{
            r_\text{src}^2 - (\vec{r}_\text{src} \cdot \uvec{n}_\pm)^2
        }
    \right) \,,
    \label{eq:arrival_time}
\end{align}
where $\vec{r}_\text{src}$ is the position of the gravitational-wave source and $\robsvec$ is the position of the observer, both with respect to the point-mass lens location.
In the second line, we used the fact that $\robsvec$ is far away, so as to write $\robsvec \cdot \uvec{n}_\pm \approx \robs$, and dropped the explicit $\bigOof{c^{-5}}$.
Taking into account the hierarchy of distances associated with the current system, the expression can actually be further simplified to remove any explicit dependence on $\Dls$ (see \cref{app:magnification_for_a_far_away_observer}).

Then,
\begin{equation}
    \Delta t_\text{arrival}
    =
    \left|
        \frac{1}{\theta_{+,\infty}^2} - \frac{1}{\theta_{-,\infty}^2}
        + \ln \left| \frac{\theta_{-,\infty}}{\theta_{+,\infty}}\right|
    \right| \,.
    \label{eq:asymptotic_arrival_time_difference}
\end{equation}

\subsection{Magnification}
\label{app:magnification_for_a_far_away_observer}

The effect of lensing and of the path bending is not only to delay the signal, but also to (de)magnify it.
For a point mass, the expression of the magnification $\mu_\pm$ of each image is well known and reads \cite{poissonGravityNewtonianPostNewtonian2014}
\begin{equation}
    \mu_\pm
    =
    \frac{\theta_\pm}{\beta} \frac{\dd \theta_\pm}{\dd \beta}
    =
    \pm \frac{1}{4}
    \left(
        \frac{\beta}{\sqrt{\beta^2+4 \thetaE^2}}
        +
        \frac{\sqrt{\beta^2 + 4 \thetaE^2}}{\beta}
        \pm
        2
    \right) \;,
\end{equation}
with $\thetaE$ defined in \cref{eq:Einstein_angle_definition}, and all other angles as defined in \cref{fig:lensing_plane_frame}.
In the present case, and in terms of the impact factor $b$, we can write
\begin{equation}
    \mu_{\pm,\infty} = \frac{
        \left(
            b \pm \sqrt{8 \Dls \RS + b^2}
        \right)^2
    }{
        4 b \sqrt{8 \Dls \RS + b^2}
    } \, .
    \label{eq:magnification_factors_for_a_far_away_observer}
\end{equation}

\section{Angle \texorpdfstring{$\thetaJL$}{θJL} between the total and the orbital angular momenta}
\label{app:angle_theta_JL}

The angle $\thetaJL$ between the orbital angular momentum $\Lvec$ and the total angular momentum $\Jvec$ is usually not considered a free parameter.
For example, it does not appear in the list of the typical $15$ gravitational-wave parameters of \cref{sec:geometrical_transformation_rules_of_angles}.
Given a reference frequency $f_\text{ref}$ at which we define a reference orbital phase $\phiorb$, we describe here how to compute the value of $\thetaJL$.

By construction, there exist a frame where the spins $s_1$ and $s_2$ of the two masses $m_1$ and $m_2$ can be expressed as
\begin{align}
    S_1 & = a_1 \, m_1^2 \cdot \begin{pmatrix}
        \sin(t_1) \\
        0 \\
        \cos(t_1)
    \end{pmatrix}
    &
    S_2 & = a_2 \, m_2^2 \cdot \begin{pmatrix}
        \sin(t_2) \cos(\phi_{12}) \\
        \sin(t_2) \sin(\phi_{12}) \\
        \cos(t_2)
    \end{pmatrix} \,,
\end{align}
where $t_1$ and $t_2$ are the tilts of the spins given with respect to the orbital angular momentum, $\phi_{12}$ is the azimuathal angle between the two spins, and $a_1$ and $a_2$ denotes their amplitudes.
We now express the orbital angular momentum $\Lvec$ explicitly as
\begin{equation}
    \Lvec = \begin{pmatrix}
        0 \\
        0 \\
        L_\text{src}
    \end{pmatrix}
    \quad \text{where} \quad
    L_\text{src}
    = \frac{m_1 \cdot m_2}{\sqrt[3]{\pi f_\text{ref} (m_1 + m_2)}} \,.
\end{equation}
Then, since $\Jvec = S_1 + S_2 + \Lvec$, the angle between $\Jvec$ and $\Lvec$ reads
\begin{equation}
    \thetaJL = \atantwo\left(
        J_x^2 + J_y^2,
        J_z
    \right) \,.
\end{equation}

\section{On the signal duration}

We present here two short calculations showing that lensing of a \ac{bbh} in an \ac{agn} is likely to yield complete and separate signals.

\subsection{Probability of truncation}
\label{app:probability_of_truncation}

As a rule of thumb, the condition for lensing to occur at all is for the emitter to be placed within the lensing paraboloid characterized by the Einstein angle $\thetaE$, \ie. $\beta \lessapprox \thetaE$.
The rightmost panel of \cref{fig:quantities_at_infinity} shows this by highlighting how the secondary signal is suddenly demagnified once $y \lessapprox 1$.
To give some quantitative insight into the question, we go through some back-of-the-envelope calculations of the chance of a signal being cut.
Firstly, the signals measured by the \ac{lvk} collaboration often last less than a second.
Knowing the orbital speed $\vorbit$ (\cref{eq:vorbit_definition}) and writing the signal duration as $\signalduration$, the arc length $\ell_\text{sig}$ travelled during this interval is simply
\begin{equation}
    \ell_\text{sig} = \vorbit \cdot \signalduration \,.
    \label{eq:signal_arc_length_definition}
\end{equation}
Secondly, the length of the total orbital trajectory that is lensed reads \cite{Leong:2024nnx}
\begin{equation}
    \ell_\text{lensed} = 2 \pi \, \rorbit \arccos\left(\frac{\sqrt{\RS^2 + \rorbit^2} - \RS}{\rorbit \, \sin(\iota_\text{AGN})}\right) \,,
    \label{eq:signal_arc_length_explicit_formula}
\end{equation}
where $\iota_\text{AGN}$ is the inclination of the accretion disk of the \ac{agn} relative to the observer.
Notice that \cref{eq:signal_arc_length_explicit_formula} implies some bounds on the pair $\{\iota_\text{AGN}, \rorbit \}$, for lensing to occur at all.
This inclination $\iota_\text{AGN}$ can be related to our angle parametrization via
\begin{subequations}
    \begin{align}
        \sin(\iota_\text{AGN}) & = \LAGNhat \cdot \Nhat \\
        &
        \begin{aligned}[t]
            {}= & -\sin (\thetaJN) (\sin (\philns) \cos (\phimis) \cos (\thetalns)+\cos (\philns) \sin (\phimis)) \\
            & + \cos (\phimis) \cos (\thetaJN) \sin (\thetalns)
            \,.
        \end{aligned}
    \end{align}
    \label{eq:iota_AGN_explicit}
\end{subequations}

Now, the chance for a lensed signal to be cut can be simply written as
\begin{equation}
    \frac{\ell_\text{sig}}{\ell_\text{lensed}} \,.
\end{equation}
Evaluating \cref{eq:signal_arc_length_explicit_formula} numerically shows that typical probabilities lie around $\num{1e-3}$ or less: it is therefore unlikely to occur for an \ac{agn} system with a supermassive black hole of mass $\MSMBH \sim \qty{1e6}{\solarmass}$ and a binary orbiting at $\rorbit \sim \qty{100}{\RSunit}$.

\subsection{Probability of signal overlap}
\label{app:probability_of_signal_overlap}

We say that the two lensed images overlap when the arrival time difference is shorter than the signal duration, \ie.
\begin{equation}
    \left|
        \Delta t_{\text{arrival},+} - \Delta t_{\text{arrival},-}
    \right|
    < \signalduration \,.
    \label{eq:condition_for_millilensing}
\end{equation}
The arrival time difference depends on the lens mass and source position (see \cref{fig:quantities_at_infinity,eq:asymptotic_arrival_time_difference}), while $\signalduration$ depends on the binary masses.
For \acp{bbh}, typical signals last \qty{1}{\second} or less, whereas arrival time differences in the parameter ranges of interest are often much longer.
A full overlap probability would require specifying the binary mass function and the distribution of source positions behind the lens; we do not pursue that integral here.

\bibliography{bibliography.bib}

\end{document}